\documentclass[article]{elsarticle}
\usepackage{lineno}
\usepackage{amsmath}
\usepackage{color}
\definecolor{lightGrey}{gray}{0.1}
\definecolor{darkGrey}{gray}{0.9}

\modulolinenumbers[1]

\usepackage[]{journals}

\usepackage[svgnames]{xcolor}
\usepackage[colorlinks]{hyperref}
\AtBeginDocument{  \hypersetup{
    citecolor=NavyBlue,
    urlcolor=blue}}

\journal{Journal of \LaTeX\ Templates}

\biboptions{authoryear}

\newcommand{\tcr}{}

\begin{document}

\begin{frontmatter}

\title{Surface Deposition of the Enceladus Plume\\and the Zenith Angle of Emissions}

\author{Ben S. Southworth\fnref{bs_fn,sk_lasp}}
\fntext[bs_fn]{University of Colorado at Boulder, Department of Applied Mathematics}

\author{Sascha Kempf\fnref{sk_phys,sk_lasp}}
\fntext[sk_phys]{University of Colorado at Boulder, Physics Department}
\fntext[sk_lasp]{University of Colorado at Boulder, Laboratory for Atmospheric and Space Physics}

\author{Joe Spitale\fnref{js_psi}}
\fntext[js_psi]{Planetary Science Institute}

\begin{abstract}
Since the discovery of an ice particle plume erupting from the 
south polar terrain on Saturn's moon Enceladus,
the geophysical mechanisms driving its activity have been the focus of substantial scientific research. The pattern and
deposition rate of plume material on Enceladus' surface is of interest because it provides valuable information 
about the dynamics of the ice particle ejection as well as the surface erosion. Surface deposition maps derived from numerical 
plume simulations by \mbox{\citet{Kempf:2010kx}} have been used by various researchers to interpret 
data obtained by various Cassini instruments. Here, an updated and detailed set of deposition maps is provided based on
a deep-source plume model \citep{Schmidt:2008cf}, for the eight ice-particle jets identified in \citet{Spitale:2007ks}, the updated
set of jets proposed in \cite{Porco:2014bk}, and a contrasting curtain-style plume proposed in \citet{Spitale:2015ka}. Methods
for computing the surface deposition are detailed, and the structure of surface deposition patterns is shown to be consistent
across changes in the production rate and size distribution of the plume. Maps are also provided of the surface deposition 
structure originating in each of the four Tiger Stripes.  Finally, the differing approaches used in \cite{Porco:2014bk} and
\citet{Spitale:2015ka} have given rise to a jets vs. curtains controversy regarding the emission structure of the Enceladus
plume. Here we simulate each, leading to new insight that, over time,  most emissions must be directed relatively orthogonal
to the surface because jets ``tilted'' significantly away from orthogonal lead to surface deposition patterns inconsistent with
 surface images. 

Data for maps are available in HDF5 format for a variety of particle sizes at \url{http://impact.colorado.edu/southworth_data}. 
\end{abstract}


\end{frontmatter}

\section{Introduction}

In 2005 the Cassini mission made the exciting discovery of a water-vapor and ice-particle plume erupting from the south
polar terrain on Saturn's icy moon Enceladus \citep{Dougherty:06a,Hansen:06a,Porco:2006ir,Spahn:2006im,Spencer:2006gt}.
Multiple Cassini traversals through the plume allowed Cassini in-situ instruments 
to collect samples of the emerging vapor \mbox{\citep{Waite:09a}} and ice particles \mbox{\citep{Postberg:2009jc}},
the larger of which likely originate from the boiling surface of the moon's subsurface ocean \citep{Postberg:2011jx}. Since then
much research has been devoted to understanding the Enceladus plume and its driving mechanism, for example, see
\citet{Brilliantov:2008be,Gao:2016,Hurford:2007ea,Schmidt:2008cf}. There is convincing evidence that the plume is by far
the strongest source of E-ring particles \mbox{\citep[for example,][]{Spahn:06b,Horanyi:09a}}
and also the dominant source of the resurfacing of Enceladus \citep[for example,][]{Jaumann:09a,Kempf:2010kx}.
However, there remain open questions about the plume, some of which may be addressed by examining surface deposits.

The purpose of this work is two-fold. First, we provide simulated surface deposition data resulting from the three primary
proposals for plume emission structure: the eight jets identified in \citet{Spitale:2007ks}, an updated set of approximately 100
sources identified in \citet{Porco:2014bk}, and a contrasting ``curtain-like'' plume proposed in \citet{Spitale:2015ka}. Multiple
particle sizes from $0.6-15$ $\mu$m are simulated for each source location, and data are generated on the impact flux in particles/sec/m$^2$
and mass deposition in mm/year across the surface of Enceladus. Initial simulated maps of surface deposition from the Enceladus
plume published in \cite{Kempf:2010kx} have received interest from the larger research community \citep[for example,][]{DiSisto:16a,Nahm:16a,
Scipioni:2017is} and, here, we provide a more complete set of maps and data with respect to source location and particle size. Using the
newly generated surface data for a curtain-style plume \citep{Spitale:2015ka} and the $\sim 100$ discrete jets proposed in
\citet{Porco:2014bk}, we provide new insight into the zenith angle of plume emissions, that is, the ``tilt'' of the jets. Specifically,
comparing simulated surface deposition patterns with the surface pattern seen in IR/UV images \citep{Schenk:2011bv} indicates that
highly tilted jets (zenith angle $\gg 15^\circ$) identified in \citet{Porco:2014bk} are not contributing substantially to surface deposition;
that is, the unique signature of highly tilted jets is not apparent in surface images. Potential reasons for this are discussed in Section
\ref{sec:angle}. The most likely explanation is that highly tilted jets experience short lifetimes and are not active long enough to
develop observable surface features.

A background on the plume model and simulations is given in Section \ref{sec:model}, along with a description of the
data. Details on computing impact flux and surface deposition can be found in the Appendix. Maps of surface deposition
as a function of time are given in Section \ref{sec:plot}. Data for surface maps are available in HDF5 format \cite{hdf5} at
\url{http://impact.colorado.edu/southworth_data}, and are summarized in the following table:
\begin{table}[h!]
\centering
{
\begin{tabular}{| c c c |}\hline
Source & \# Locations & \# Particle sizes \\ \hline
\citet{Spitale:2007ks} & 9 & 12 \\
\citet{Porco:2014bk} & 98 & 7 \\
\citet{Spitale:2015ka} & 115 & 7 \\\hline
\end{tabular}
\caption{Impact flux is available in HDF5 files as described here. Each HDF5 file corresponds to a given source location,
with dataspaces for each particle size, and attributes describing source direction and location, source opening angle, particle size,
number of particles launched, and number of particles collided.}
}
\end{table}

Section \ref{sec:angle} introduces the jets vs. curtains
controversy and provides evidence that, regardless of whether emissions originate from discrete jets or in a continuous curtain-style
emission, the zenith angle of emissions is largely close to orthogonal to the surface. Implications and other open questions that
surface deposition may provide insight towards are discussed in Section \ref{sec:conc}.

\section{Plume model}\label{sec:model}

Here we assume that the Enceladus plume is fed by a ``deep-source'' mechanism \citep{Brilliantov:2008be,Schmidt:2008cf,Postberg:2011jx},
where fractures in Enceladus' icy crust extend down to a liquid-water reservoir. Particles then condense and are accelerated
though a back-pressurized gas flow exiting the fracture, for which the particle velocity upon ejection takes the following distribution\footnote{Eq.
\ref{eq:Pvr} includes a correction of $1/v_{gas}$ that was omitted in \citet{Schmidt:2008cf}. That correction also appeared
without comment in \citet{Southworth:2015im}.}
\begin{equation}
	p(v | r)  = \left(1+\frac{r}{r_{c}}\right) \\
		    \frac{r}{r_{c}} \frac{v}{v_{gas}^2} \\
		    \left(1 - \frac{v}{v_{gas}}\right)^{\frac{r}{r_{c}} - 1}, \\
	\label{eq:Pvr} \end{equation}
where
\begin{equation}
	\int_0^{v_{gas}} p(v|r) dv = 1,		\label{eq:Pvr_norm} 
\end{equation}
The velocity distribution (Equation (\ref{eq:Pvr})) assumes that particle velocities cannot
be larger than the gas velocity, $v_{gas}$,  hence the normalization integral in Equation (\ref{eq:Pvr_norm}) over
$[0,v_{gas}]$.\footnote{{Note that, for this model, the particle velocity upon emission is effectively determined by the depth of its final
collision with a fracture wall before emission. Because the expected mean free particle path is on the order of decimeters \citep{Schmidt:2008cf},
fractures need not be ``deep'' for these equations to hold, as long as the driving physics remains consistent.}}
Evidence of a deep-source plume mechanism can be found in \citet{Schmidt:2008cf,Postberg:2011jx} and \citet{Yeoh:2015hg}.
In Equation (\ref{eq:Pvr}), $v_{gas}$ is the gas velocity, and $r_c$ the so-called critical radius,
which is effectively a measure of the length of time a particle has to be reaccelerated by the gas
between its final collision with a fracture wall and ejection. Particles $r < r_c$ are
efficiently accelerated to velocities approaching $v_{gas}$, while particles $r >r_c$
move in the gas flow at average velocities less than $v_{gas}$. A detailed look at the
critical radius, $r_c$, and gas velocity, $v_{gas}$, can be found in \citet{Schmidt:2008cf} and \citet{Southworth:2015im}.

\tcr{In the detailed model of plume-particle speed distribution, derived in \citet{Schmidt:2008cf}, $r_c$ and $v_{gas}$
are actually nonlinearly coupled variables. To that end, simulations of the venting process were run in \citet{Schmidt:2008cf}
to produce a discrete probability distribution over a set of particle radii, rather than an analytical distribution with fixed
$r_c$ and $v_{gas}$, as in Equation \eqref{eq:Pvr}. For simulations of full jet- and curtain-models performed here, the
discrete speed distribution developed in \citet{Schmidt:2008cf} is used to weight particle velocities. The parameter space of $v_{gas}$
and $r_c$ is also explored in Section \ref{sec:plot} by applying an analytic distribution of the form in Equation \eqref{eq:Pvr},
with fixed values of $r_c$ and $v_{gas}$,
to simulations of the eight sources in \citet{Spitale:2007ks}. Note that for parameter values $r_c \approx 0.2$
$\mu$m and $v_{gas} \approx 700$ m/s, the analytic speed distribution in Equation \eqref{eq:Pvr} is close to the
discrete speed distribution resulting from simulations of the venting process \citep{Schmidt:2008cf}.}

The size-dependent speed distribution is consistent with a chemically stratified plume, as evidenced by data from the Cassini
Cosmic Dust Analyzer (CDA) \citep{Postberg:2011jx}, as well as surface deposition patterns that depend on particle size
\citep{Kempf:2010kx, Scipioni:2017is}. Particle ejection angles are assumed to be azimuthally uniform and follow a
cos$^2(\theta)$-zenith angle distribution over $\theta$ between $0^{^\circ}$ and $15^{^\circ}$. A maximum half-angle of
$15^{^\circ}$ is consistent with opening angles seen in \citet{Spitale:2015ka}, and the $\cos^2$-distribution indicative of the smooth
onset, peak and decline of particle impact rates as seen by CDA \citep{Kempf:2010kx}. A plume source is simulated by launching millions of
particles from a given location and integrating their trajectories in a Saturn-centered quasi-inertial frame until each particle
has either collided with Enceladus, or escaped from Enceladus and established orbit about Saturn. The equations of motion
account for Saturn's gravity, Enceladus' gravity, and electromagnetic forces, including particle charging \citep{Horanyi:1996fl},
in a Z3-Voyager magnetic field about Saturn \citep{Connerney:1993fp}.
{We have also implemented a magnetic field based on a local interaction model between plasma and the Enceladus plume,}
as proposed in \citet{Simon:2011ef}, which considers the
effects of the Enceladus plume on the corotating plasma in Saturn's magnetosphere. Although the local model in \citet{Simon:2011ef}
reproduces data from the Cassini magnetometer (MAG) instrument more faithfully than a global magnetic field about Saturn,
overall plume dynamics for the particle sizes considered here ($>0.6$ $\mu$m) are nearly identical using a Z3-charging model, a
local charging model, and no particle charging.  In particular, surface deposition patterns are not affected by a change in the charging
equations considered. Further details on the software used to run simulations as well as the equations of motion and underlying
distributions can be found in \citet{Schmidt:2008cf,Kempf:2010kx,Southworth:2015im}. {In particular, the Appendix of \cite{mass}
provides a detailed description of all aspects of the software and modeling techniques.}

Particle sizes between $0.6-15$ $\mu$m are simulated for each source location, leading to $10^6-10^{7}$ particle simulations per
source. Twelve sizes are simulated for the eight jets identified in \citet{Spitale:2007ks}, and seven sizes simulated for the curtain
model \citep{Spitale:2015ka} and updated 100 jets proposed in \citet{Porco:2014bk}.
Particle trajectories are integrated until either {the particle completes two orbits about Saturn without entering Enceladus' Hill sphere,
or collides with the surface of Enceladus.} When a particle collides with Enceladus, its position and velocity at the time of collision are saved with
respect to an Enceladus-centered inertial frame (these data are available on request). All collisions for a given particle size and
source location are then grouped into $1^\circ$-latitude $\times$ $1^\circ$-longitude bins, covering the surface of Enceladus.
At the meridian, one bin covers an approximate square with dimensions $4.35$ km $\times 4.35$ km and a surface area of
approximately 19 km$^2$; at the poles, one bin covers a surface area of approximately 0.17 km$^2$. Bins are then normalized
to give the contribution of a single ejected plume particle to impact rate per m$^2$ in each bin. Data for each simulated particle
size and jet location are stored in $360\times 180$ arrays, corresponding to planetographic coordinates in western longitude.
Scaling the impact flux for each bin by the size of the bin, and summing over the entire array gives the fraction of simulated
particles that collided with the surface. For particles larger than 1 $\mu$m, this is close to one (that is, most particles larger than
$1\mu$ do not escape Enceladus' gravity). These arrays are available in HDF5 format \citep{hdf5} at
\url{http://impact.colorado.edu/southworth_data}. 

Details on computing the impact flux and surface deposition can be found in the Appendix. It is assumed
that particles follow a power-law size distribution with slope $\alpha$, $p(r) \sim r^{-\alpha}$, and the plume has a total mass production
rate $M^+$ kg/sec. Values for the size-distribution slope and mass production are based on CDA data \cite{Kempf:2010kx,mass}, but
surface deposition patterns are shown to be consistent across changes in $\alpha$ and $M^+$ (Section \ref{sec:plot}). In any case, because
data arrays are stored for individual particle sizes, deposition and impact flux can be re-weighted with arbitrary size distribution models
and mass production.

\begin{figure}[!b]
\centering
\includegraphics[width=\textwidth]{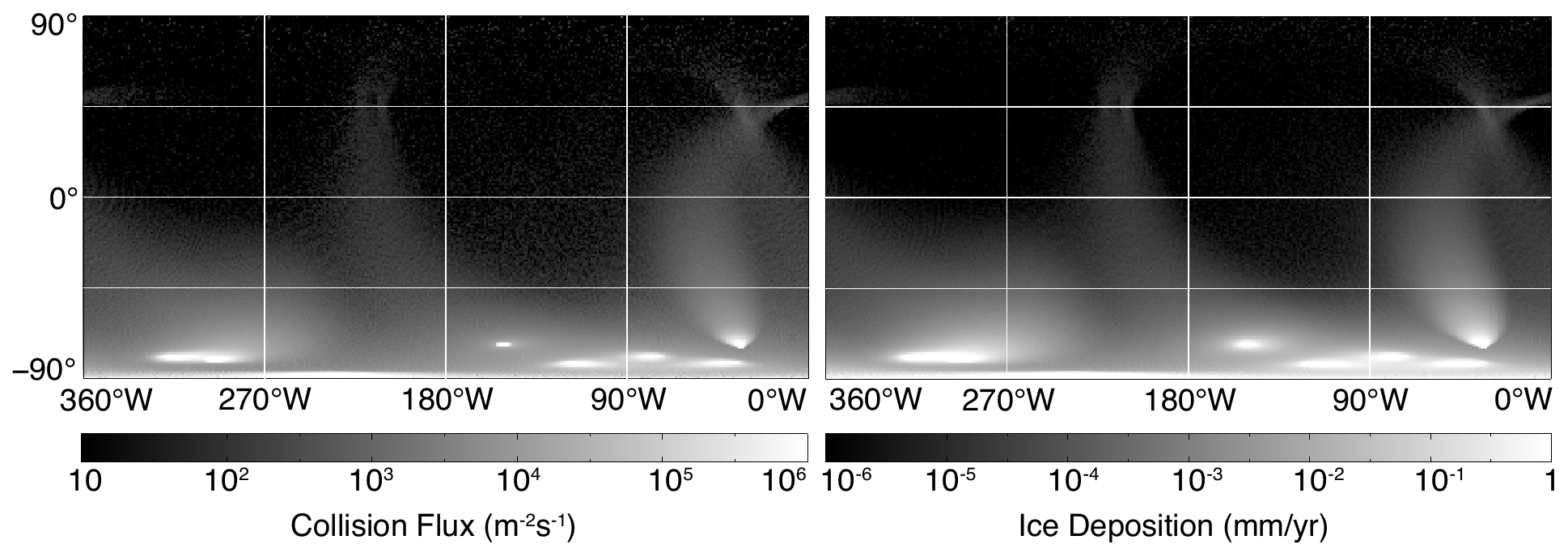}
\caption{Cumulative plume particle deposition on Enceladus' surface in mm/year for the eight sources proposed in \citet{Spitale:2007ks},
particle sizes $0.6-15$ $\mu$m, assuming a mass production rate of $M^+=25$ kg/s, and slope of the power-law size distribution $\alpha=3.1$.}
\label{fig:dep_mass}
\end{figure}

Source locations consist of the eight sources initially identified in \cite{Spitale:2007ks} and the updated 98 sources
identified in \citet{Porco:2014bk}, all simulated as published, with a given direction, and relative source strength
based on the number of sightings \citep{Spitale:2007ks,Porco:2014bk}. A curtain is simulated
via discrete sources spaced evenly along the tiger stripes, approximately 5 km apart, directed orthogonal to the surface. {Although
this is not a true ``curtain,'' simple calculation shows that emissions with an opening angle of 15$^\circ$ reach out approximately 2.5km laterally at an
altitude of 9km and approximately 5km laterally by 18km in altitude. Thus, at the Cassini flyby altitudes of 50km and higher, emissions from
these 5km-spaced sources will appear as a curtain, and the discrete nature of sources offers greater numerical flexibility in adjusting
the production rate along fractures a posteriori.} Relative production rate of each source in the curtain is based on the average activity
of emissions along the Tiger Stripes, measured through images of the plume in \citet{Spitale:2015ka}. {In the model proposed by
\citet{Porco:2014bk}, some jets are closer than 5km in proximity, and others much more spread out. The fundamental difference
between the two models that comes up in this study is the direction that emissions are pointing with respect to the surface.}

\section{Surface maps}\label{sec:plot}

This section provides maps of surface deposition for various particle sizes and model parameters. It is important to note that our
simulated deposition depth assumes perfect compaction of particles, and particle density, $\rho_p$, equal to that of water ice. {However, as
can be seen in the Appendix (Equations \eqref{eq:imp2} and \eqref{eq:h2}), the impact flux and mass deposition scale with a factor of $1/\rho$.
For global surface deposition, the density $\rho$ represents an average density of deposited material. More generally, if particles deposit with
some porosity $\psi$, then the average deposit density will scale like $\rho = (1-\psi)\rho_p$, where $\rho_p$ is the average individual
particle density. Deposition and impact flux for non-perfect compaction, $\psi > 0$, can be obtained by scaling results presented here by
$\frac{1}{1-\psi}$. In fact, evidence for ``fluffy'' (less dense) particles \citep{Gao:2016} as well as the fact that particles are very
unlikely to pack perfectly on the surface, suggests that deposition is almost certainly greater than presented here. Nevertheless,
assuming perfect compaction and water-ice-density provides a lower bound on deposition.}

\begin{figure}[!b]
\centering
\includegraphics[width=\textwidth]{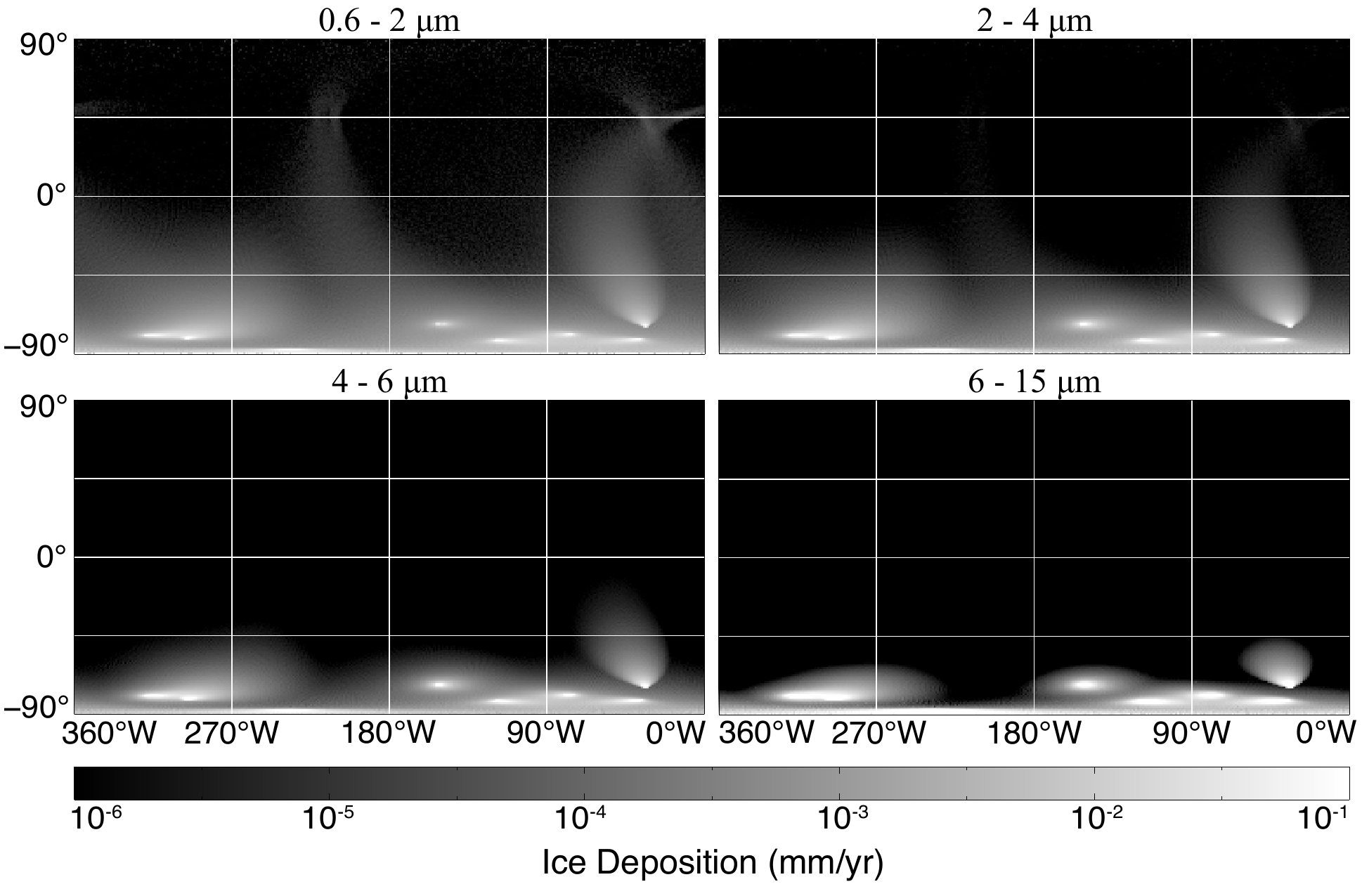}
\caption{Plume particle deposition rates on Enceladus' surface in mm/year for the eight sources proposed in \citet{Spitale:2007ks} and for
particle sizes $0.6-2$ $\mu$m, $2-4$ $\mu$m, $4-6$ $\mu$m, and $6-15$ $\mu$m. Mass production rate is $M^+=25$ kg/s, the slope of the
power-law size distribution is $\alpha=3.1$, and the total particle size range in terms of mass production is $0.6-15$ $\mu$m. Combining these four plots corresponds to the total deposition rate for particles of size $0.6-15$ $\mu$m. }
\label{fig:dep_large}
\end{figure}

\begin{figure}[!t]
\centering
\includegraphics[width=\textwidth]{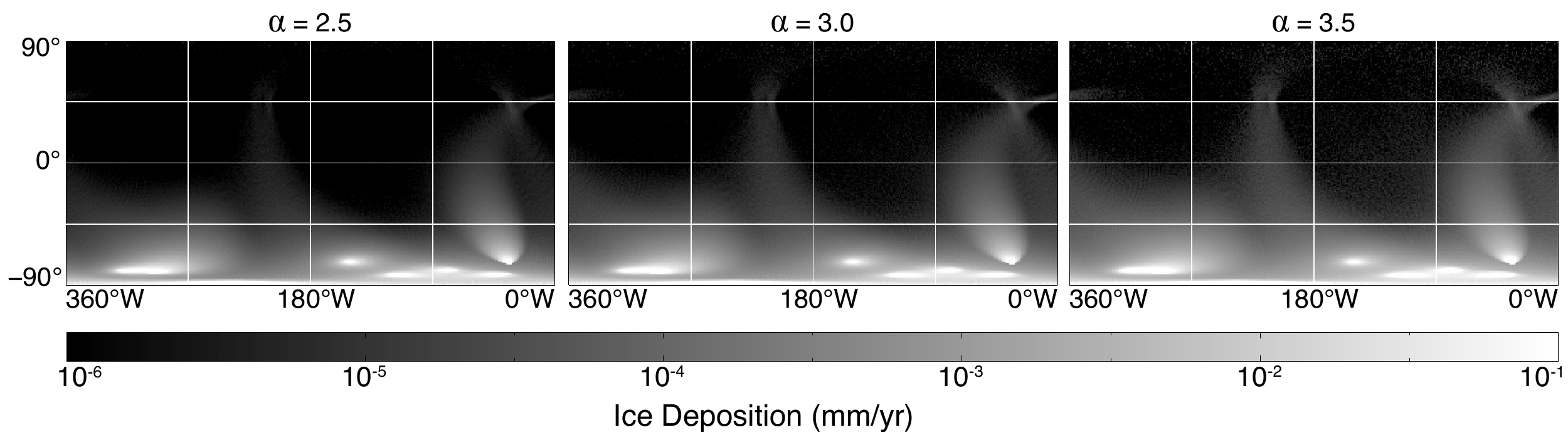}
\caption{Cumulative plume particle deposition on Enceladus' surface in mm/year for the eight sources proposed in \citet{Spitale:2007ks},
particle sizes $0.6-15$ $\mu$m, assuming a mass production rate of $M^+=25$ kg/s, and varying the slope of the power-law size distribution,
$\alpha$. }
\label{fig:dep_alpha}
\end{figure}

\begin{figure}[t!]
\centering
\includegraphics[width=\textwidth]{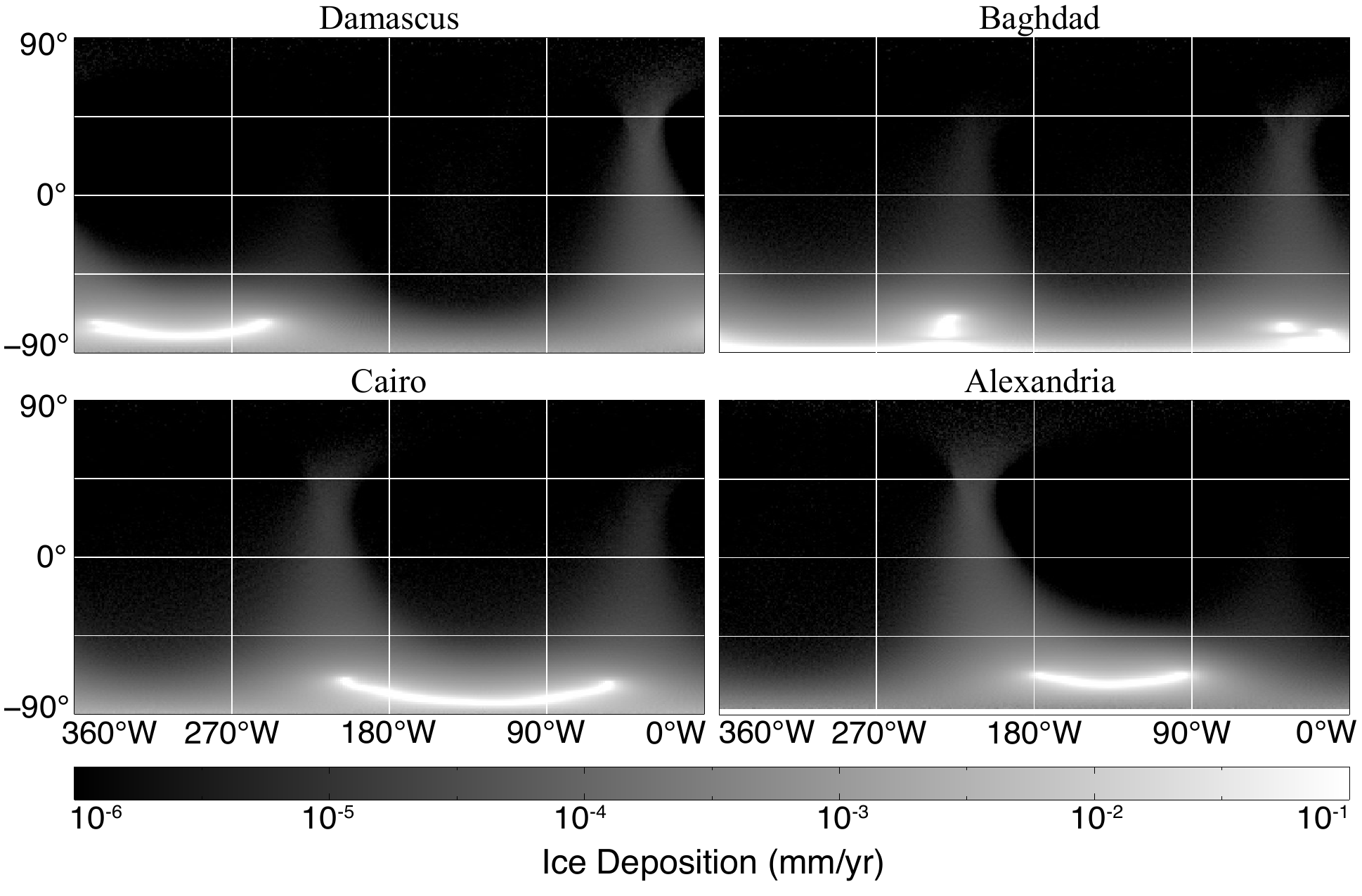}
\caption{Plume particle deposition rates on Enceladus' surface in mm/year for particle sizes $0.6-15$ $\mu$m
and all particles emitted in a curtain-style plume \citep{Spitale:2015ka} from a single fracture. Mass production rate is $M^+=25$ kg/s for each
fracture and the slope of the power-law size distribution is $\alpha=3.1$. Note, these images do not account for potentially different
production rates from different fractures. The fractures can be seen as the bright curves in each map where deposition rate is highest. }
\label{fig:fracture}
\end{figure}

Figures \ref{fig:dep_mass}, \ref{fig:dep_large}, and \ref{fig:dep_alpha} are based on the original eight sources identified in \citet{Spitale:2007ks},
and Figure \ref{fig:fracture} is based on a continuous curtain emission over each individual fracture. Maps for jet sources proposed
in \citet{Porco:2014bk} and a full curtain scenario as proposed in \citet{Spitale:2015ka} are given in Section \ref{sec:angle}.

Figure \ref{fig:dep_mass} maps the particle impact flux in particles/sec/m$^2$ and surface deposition in mm/year for particles
$0.6-15$ $\mu$m, and Figure \ref{fig:dep_large} breaks the total deposition down into particle size ranges starting at
$0.6-2.0$ $\mu$m, and increasing up to $6-15$ $\mu$m.\footnote{The last bin contains a large range of sizes because there are very few particles
of that size, and they do not travel far from the fractures} For each of these figures, we choose $r_{min} = 0.6$ $\mu$m
and $r_{max} = 15$ $\mu$m, a mass production rate of $M^+ = 25$ kg/s, and a size distribution slope of
$\alpha = 3.1$, based on fitting impact rates to reproduce Cassini CDA flyby data \citep{mass}. All results presented use
these parameters, unless stated otherwise. 

The mass production and size distribution slope are directly motivated through CDA data on low-altitude flybys \citep{mass},
which provide the most direct measurements to date of large particles in the Enceladus plume. Results are also relatively
consistent with estimates of mass production in \citet{Meier:2015ba,Porco:2017kh}. Note that in
fixing $r_{min}$ and $r_{max}$, $M^+$ corresponds to the mass production of particles \textit{in this size range}. A maximum particle size
must be chosen so that the average mass of a particle is well-defined; here we choose $r_{max}$ to be sufficiently large that plume
particles of that size or larger are very unlikely and increasing $r_{max}$ has a small effect on results. 
We only consider micron-size particles (formed through condensation in gas flow)
because the assumed power-law size distribution does not necessarily propagate back to nano-grains. Nano-grains
can also be formed through supersonic nucleation bursts at the narrowest channel points, which can occur at various depths, leading to
a bi-modal or multi-modal size distribution. In any case, it is easily verified that the total ejected mass and the total redeposition
mass are dominated by large particles, so extending $r_{min}$ to nano-grains also does not have a significant
effect on results (see Supplementary Material in \citet{Schmidt:2008cf}). Although there is evidence that plume strength varies over time
\citep{Hedman:2013gf,Nimmo:2014ks,mass}, in considering
mass deposition on the surface, we need only consider an average mass production. 

In Equation \ref{eq:h2} of the appendix, we show that mass deposition depends linearly on the mass production
rate and, thus, the structure of surface deposition is consistent across changes in $M^+$. 
Similarly, Figure \ref{fig:dep_alpha} shows that the structure of the surface deposition pattern is not strongly affected by changes
in size distribution slope, $\alpha$, either. A steeper size distribution (larger $\alpha$; see $\alpha=3.5$, Figure \ref{fig:dep_alpha})
results in more small particles ejected, which tend to have higher ejection velocities \citep{Schmidt:2008cf,Hedman:2009ka} and
travel larger distances from the source, leading to an increase in deposition away from the south pole. Conversely, a flatter
size distribution slope (smaller $\alpha$; see $\alpha=2.5$, Figure \ref{fig:dep_alpha}) results in more large particles ejected,
which have slower initial velocities and impact the moon closer to the source location. Then, there are increased deposition
rates close to the source, and decreased deposition rates far from the source. In any case, the general pattern of plume
resurfacing seen in Figures \ref{fig:dep_mass}, \ref{fig:dep_large}, and \ref{fig:dep_alpha} is similar for all values of $\alpha$
and $M^+$.

\begin{figure}[!t]
\centering
\includegraphics[width=\textwidth]{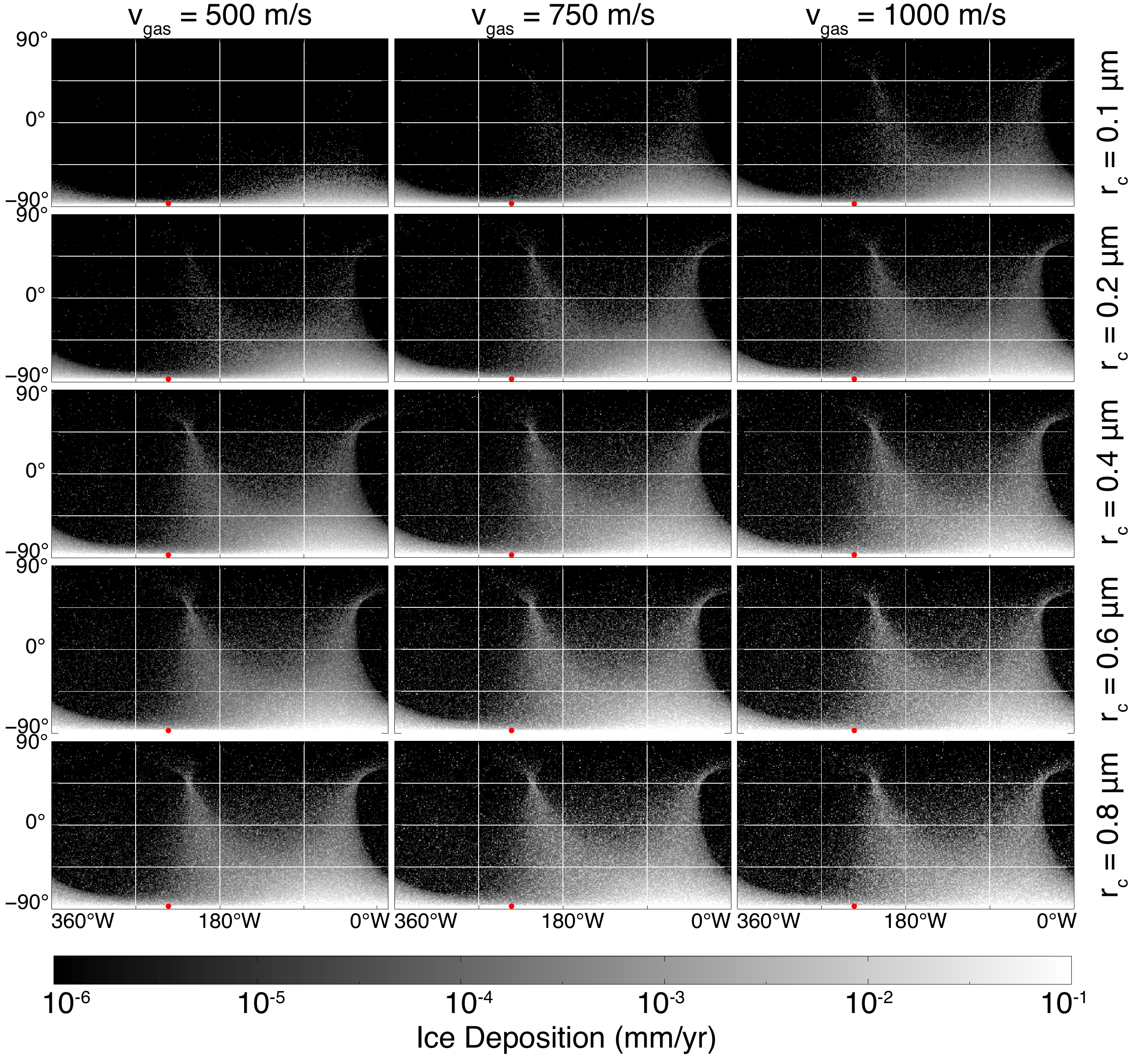}
\caption{{Cumulative plume particle deposition on Enceladus' surface in mm/year for the B2 jet \citep{Spitale:2007ks} (location
marked with red dot), with particle sizes $0.6-15$ $\mu$m, assuming a mass production rate of $M^+=25$ kg/s and power-law
size distribution  slope of $\alpha=3.1$. Results are shown for three gas velocities and four critical radii.}}
\label{fig:speeddist}
\end{figure}

{For most results, we fix a speed distribution based on \citet{Schmidt:2008cf}, which approximately correlates with parameters
$r_c = 0.2$ $\mu$m and $v_{gas} = 700$ m/s. However, Figure \ref{fig:speeddist} presents estimated mass production rates for the
B2 jet identified in \citet{Spitale:2007ks}, with twelve combinations of $r_c$ and $v_{gas}$, encompassing a wide range of possible
plume configurations. Note that, for the most part, the structure of deposition does not change based on a modified speed distribution.
Only the deposition rate changes, generally increasing with higher gas velocity and critical radius, both of which contribute to large
particles, which dominate mass deposition, traveling farther from their emission location. The one exception is the case of a low gas
velocity and small critical radius, whereby large particles do not make it sufficiently far from the source, and deposition features
are confined to a smaller region close to the emission source.}

So far we have only considered the eight sources published in \citet{Spitale:2007ks}. Figure \ref{fig:fracture} shows the
deposition from a curtain-style emission (see Section \ref{sec:angle:background}) isolated to the four main Tiger Stripes
of the Enceladus plume. Due to three-body effects and the angle of ejection associated with the surface normal, emissions from
the outer-most fractures, Alexandria and Damascus, are most likely to reach the north polar region and would likely dominate resurfacing there. 
Here we have assumed emissions are directed orthogonal to the surface; note that highly-tilted emissions from Baghdad
or Cairo may also be likely to reach the north pole, but in Section \ref{sec:angle} we discuss that such emissions are
generally not active for long periods of time. Further images of deposition from a curtain-style plume \citep{Spitale:2015ka}
and the jets proposed in \citet{Porco:2014bk} can be found in Section \ref{sec:angle}. 

\section{Jets vs. curtains and tilting of emissions}\label{sec:angle}

\subsection{Competing theories of emission structure}\label{sec:angle:background}

The structure and location of plume emissions on Enceladus' surface was first studied by triangulating images of observed
jetting activity from various angles, and projecting the result back to an approximate source location \citep{Spitale:2007ks}.
\citet{Spitale:2007ks} identified eight distinct jet sources that are largely consistent with thermal emission signatures measured by 
the Cassini Composite Infrared Spectrometer (CIRS) \citep{Spencer:2006gt}; however, the image resolution was relatively coarse
and the accuracy of proposed source locations no better than 10 to 20 km. This led to a set of follow-up observations at closer ranges
to better resolve the emission structure of the plume. 
\citet{Porco:2014bk} analyzed six years of imaging data from the Cassini Imaging Science Subsystem (ISS) using a triangulation-based
approach, resulting in the identification of approximately 100 discrete ``jets.'' Results from \citet{Porco:2014bk} are consistent with
temperatures measured across the south polar terrain by CIRS \citep{Howett:2011gw} as well as localized hot spots
identified in Cassini Visible and Infrared Mapping Spectrometer (VIMS) observations \citep{Goguen:2013ee}.

More recently, \citet{Spitale:2015ka} noticed that much of the plume activity actually appears as a relatively continuous glow in ISS images, as
opposed to discrete jet-like features as proposed in \citet{Porco:2014bk}, and that finer structure within the plume is difficult to reliably
identify over successive ISS images. This motivated a different analysis applied to many of the same data sets used in \citet{Porco:2014bk},
where a continuous ``curtain'' emission is simulated over fractures and compared with images of the plume to identify active regions.
One result that came out of that study is that so-called ``phantom-jets'' may appear in an image of a sinuous fracture, corresponding to
regions where continuous curtain emissions overlap in the line of sight and (falsely) appear as discrete jetting activity. Although
many jets identified in \citet{Porco:2014bk} are undoubtedly real and have shown to be consistent with other data
\citep{Howett:2011gw,Helfenstein:2015ja}, it is likely that some of the jets identified in \citet{Porco:2014bk} are phantom jets.
The controversy was also recently addressed in \citet{teolis2017enceladus}.

The differing approaches and results of \citet{Porco:2014bk} and \citet{Spitale:2015ka} have stimulated an in-depth review
of the interpretation of ISS images and plume emission structure. Here, we simulate the jets proposed in \citet{Porco:2014bk}
as well as an approximate curtain, consistent with \citet{Spitale:2015ka}. Although the results presented here do not favor either
approach, a comparison with surface color maps \citep{Schenk:2011bv} does provide constraints on the zenith angle or ``tilt'' of
emissions relative to the surface.

\subsection{Surface deposition and highly-tilted emissions}\label{sec:angle:background}

\begin{figure}[!h]
\begin{center}
\includegraphics[width=\textwidth]{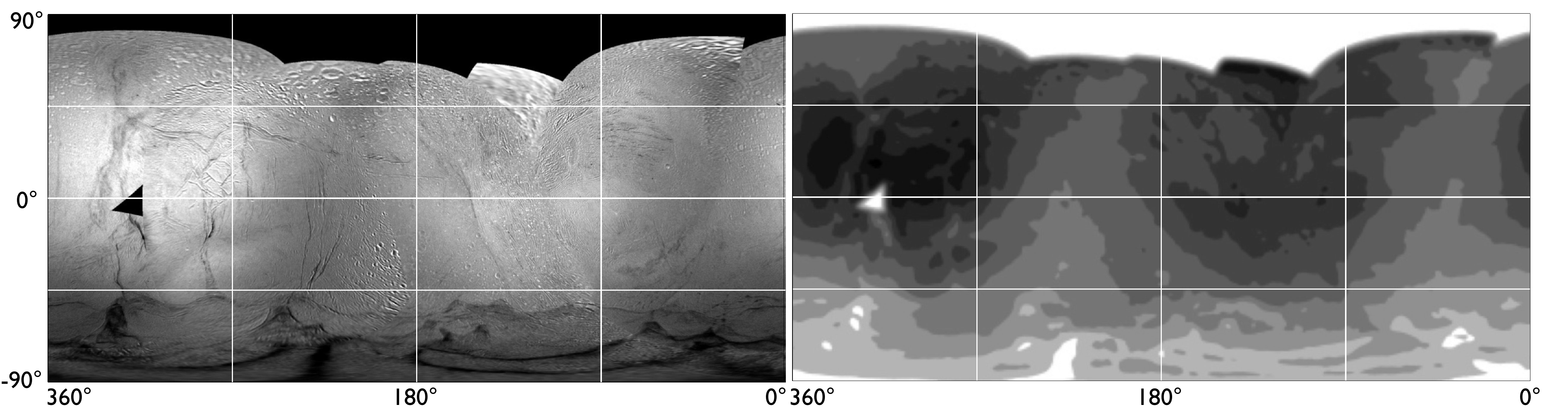}
\caption{IR/UV color maps of Enceladus, as provided in \citet{Schenk:2011bv} on the left, along with
a log of the IR/UV ratio to compare with simulated deposition rates in log-scale on the right. Darker areas in the
IR/UV ratio (left) correspond to surface area with reflectivity similar to that of unaltered water-ice, which
is believed due to a resurfacing effect of Enceladus by plume particles from a subsurface ocean. Note that
in the log-scale (right) the contrast reverses and such areas appear lighter. 
}
\label{fig:surface}
\end{center}
\end{figure}

In \citet{Schenk:2011bv}, near-global, high-resolution color maps of Enceladus were constructed using data from the Cassini ISS in
three colors, UV, Green, and near-IR. Looking at the IR/UV ratio provides a color contrast, where a ``reddish'' area on the surface
appears bright, and a ``blueish''  area, potentially corresponding to unaltered water ice, appears darker [\citep{Schenk:2011bv}; FIgure \ref{fig:surface}, left panel].
It is generally believed that Enceladus' unique color pattern, differing from other Saturnian satellites, is a result of surface re-deposition
due to plume activity \citep{Hendrix:2010gq,Kempf:2010kx,Schenk:2011bv}, which agrees with comparisons of surface maps and
simulated deposition patterns. Here, we assume that the surface pattern seen on Enceladus and shown in Figure \ref{fig:surface}
does indeed result from plume deposition and use this as a basis for expected surface deposition patterns in simulated plumes. 

The Enceladus plume model and structure can be constrained by ensuring that data collected by spacecraft are reproducible.
Here, we use IR/UV maps (Figure \ref{fig:surface}) as a reference to compare simulated surface deposition profiles. 
Figure \ref{fig:dep} shows simulated global surface deposition profiles of plume particles size $0.6-15$ $\mu$m, for both a curtain-style
plume \citep{Spitale:2015ka} and discrete jet sources \citep{Porco:2014bk}. Each simulated plume leads to a surface deposition
pattern that is largely consistent with the IR/UV ratio seen in surface images of Enceladus (Figure \ref{fig:surface}). \tcr{However, 
the discrete jets proposed in \citet{Porco:2014bk} lead to a number of surface features and patterns that are not seen in surface
images, while the curtain-style plume does not. Some of the features produced by the jets model include diagonal bands centered
near $(290^\circ$W, $30^\circ$S) and ($330^\circ$W, $0^\circ$S), fine-scale structure near ($180^\circ$W, $45^\circ$N), a wider band
near $(135^\circ$W, $30^\circ$S), and the isolated spot near ($100^\circ$W, $40^\circ$N). Although
a rigorous model connecting deposition rates to surface reflectivity is an open topic, simulated surface features that \textit{cannot} be
seen in imaging data indicates that some of the simulated jets are not actually depositing significant material on the surface. 
Conversely, the curtain model does not lead to resurfacing in certain areas which images indicate have been resurfaced,
some of which the jet model is able to reproduce, particularly terrain near the north pole, such as ($350^\circ$W, $70^\circ$N)
and ($15^\circ$W, $50^\circ$N). The terrain in approximate areas ($90^\circ$W--$180^\circ$W, $0^\circ$S--$45^\circ$S) and
($270^\circ$W--$360^\circ$W, $0^\circ$S--$45^\circ$S) does not match either model perfectly; images indicate more deposition
than seen in the curtain model, but deposition patterns do not match the jet model. 
In fact, understanding why the jet model leads to features not seen in surface
images also explains why the curtain model does not reproduce some features. The remainder of this section discusses
which jets from \citet{Porco:2014bk} lead to features inconsistent with surface images, and possible explanations for why these
features are not seen in imaging data.

\begin{figure}[!h]
\begin{center}
\includegraphics[width=\textwidth]{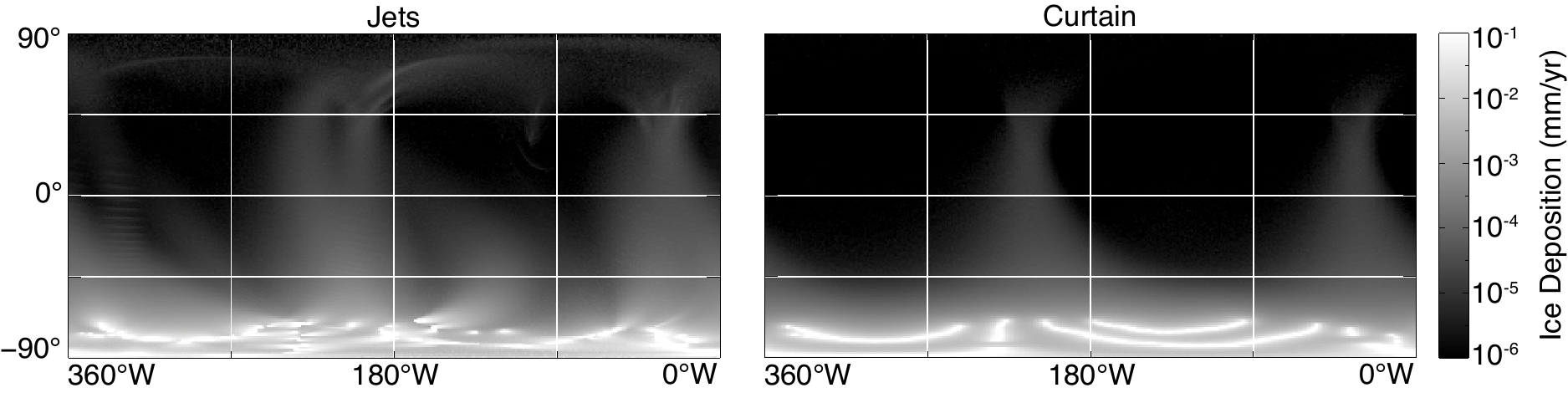}
\caption{Surface deposition rates in mm/year for jets from \citet{Porco:2014bk} on the left,
and a curtain \citep{Spitale:2015ka} on the right. Particle size ranges considered are
$0.6-15$ $\mu$m, with parameters $M^+ = 25$ kg/s and $\alpha = 3.1$.
}
\label{fig:dep}
\end{center}
\end{figure}

In looking at the simulated curtain- and jet-style plumes, the fundamental difference between the two is the
direction that jets are pointing. Like the jets, the curtain is also ``discrete'' and not simulated as a truly
continuous curtain, and both models have emissions primarily aligned on the Tiger Stripes. However, all
sources simulated for the curtain are directed orthogonal to the surface, while each of the jets
proposed in \citet{Porco:2014bk} have a given zenith and azimuthal angle. A number of the proposed zenith
angles are as large as 30--62$^\circ$ (measured from orthogonal to the surface). Such strongly tilted
jets lead to very distinct deposition patterns on the surface, which do not always agree with observed deposition.
As an example, Jets 23 and 95 from \citet{Porco:2014bk} originate in close proximity to each
other, but Jet 23 has a near-orthogonal zenith angle of $3^\circ$, while Jet 95 has a large zenith of
$42^\circ$. Figure \ref{fig:comp23_95} compares the deposition pattern for Jets 23 and 95.

\begin{figure}[!th]
\begin{center}
\includegraphics[width=\textwidth]{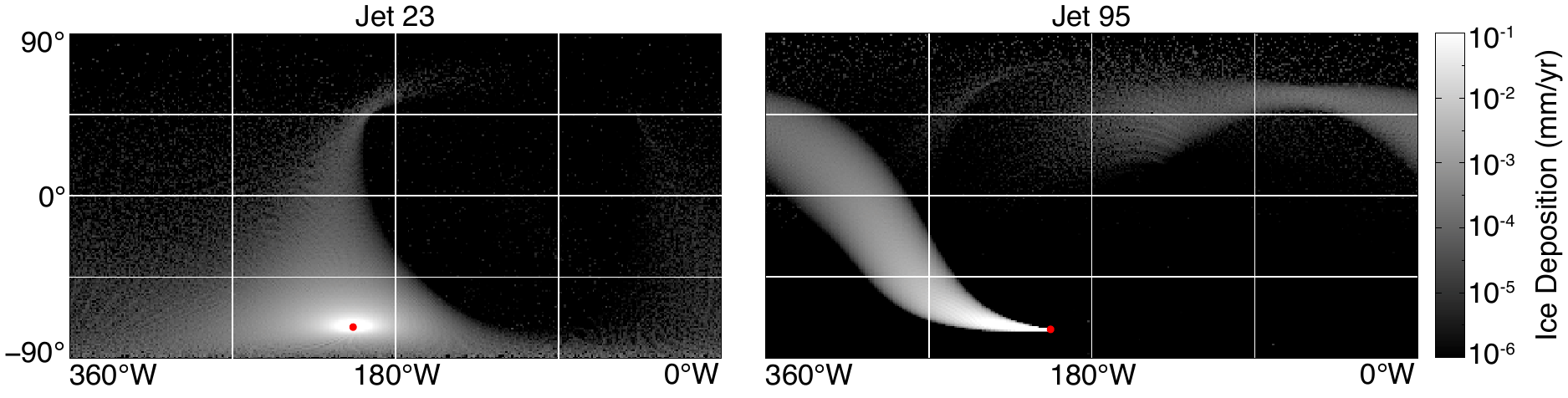}
\caption{Surface deposition rates in mm/year for jets 23 and 95 from \citet{Porco:2014bk} (marked with red dots) with particle sizes
$0.6-15$ $\mu$m, size-distribution slope $\alpha = 3.1$, and all mass production ($M^+ = 25$ kg/s) allocated to the single jet.
The contribution in mm/year with respect to all 98 jets proposed in \citet{Porco:2014bk} is approximately two orders of magnitude
smaller. }
\label{fig:comp23_95}
\end{center}
\end{figure}

Surface deposition from Jet 23 is consistent with surface IR/UV maps, but Jet 95 has a long, narrow
deposition pattern, aligned effectively the opposite direction as the pattern seen in IR/UV maps. 
Although this is partially due to the proposed azimuthal angle as well, in fact, most highly tilted jets
lead to deposition patterns that are at odds with the observed IR/UV maps. A natural conclusion from this
is that highly tilted jets do not make a major contribution to surface deposits. Figure \ref{fig:zenith} shows
the jet-plume surface deposition for three scenarios:
all jets in \citet{Porco:2014bk}, jets with zenith angle less than $30^\circ$, and jets with zenith angle less
than $20^\circ$. We can see that by simply removing highly-tilted jets, the surface deposition pattern
becomes consistent with that of the simulated curtain and, more importantly, IR/UV images of the surface.

\begin{figure}[!htb]
\begin{center}
\includegraphics[width=\textwidth]{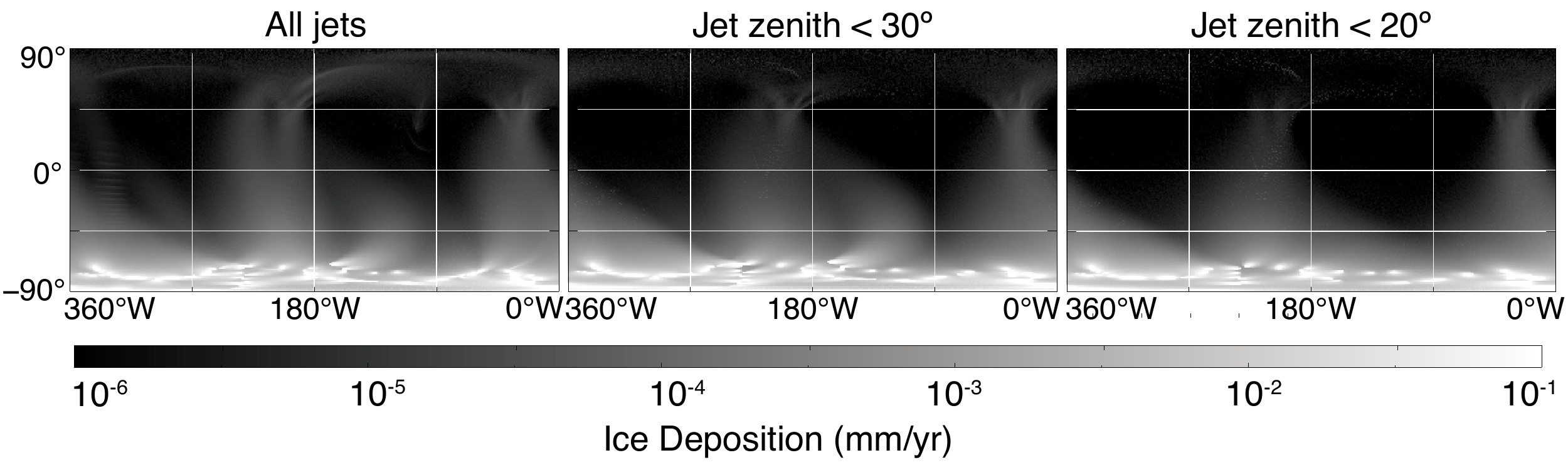}
\caption{Surface deposition rates in mm/year for jets from \citet{Porco:2014bk} and particle sizes $0.6-15$ $\mu$m,
with parameters $\alpha = 3.1$ and $M^+ = 25$ kg/s.
The leftmost figure corresponds to the 98 principle jets proposed in \citet{Porco:2014bk}, the center figure to the
86 jets with zenith angle less than $30^\circ$, and the rightmost figure to the 70 jets with zenith angles less than $20^\circ$.}
\label{fig:zenith}
\end{center}
\end{figure}

There are a number of possible explanations as to why highly-tilted emissions are not contributing to surface deposition.
The first is that many of the highly tilted jets, which are the source of surface patterns that differ from observed IR/UV ratios,
are phantoms. 
Although surface deposition of tilted jets is largely inconsistent with observed IR/UV ratios, the distinct angle
of such jets, differing from most plume emissions, does decrease the probability of these being phantom jets
\citep{Spitale:2015ka}. The tilted jets also tend to be highly prominent and consistent in images, further arguing against
a phantom origin. 

{Another possibility is that speed or size distributions may differ across emission locations, and the highly tilted jets are prone
to distributions that do not lead to significant surface deposition. As seen in Figure \ref{fig:speeddist}, in particular, a small
critical radius can limit the extent to which deposition occurs because large particles, which dominate surface deposition,
do not achieve high initial velocities. The critical radius is effectively a measure of the depth of the final collision of a particle
with a fracture wall before ejection, and, in a modeling sense, the average mean free path of the particle. It is plausible that
particles flowing through fractures with a large zenith angle of emission collide with the fracture wall closer to the moon's
surface, due to the changing angle of the fracture. Then, particles have less time to be accelerated by the gas, and even
a high gas velocity does not efficiently reaccelerate large particles, limiting the distance they can travel from the source.
However, this is again inconsistent with the prominence of highly tilted jets observed in \citet{Porco:2014bk}.
}

Finally, the most likely explanation is that highly tilted jets are just not active long enough (at least for a fixed direction of emission)
to create visible surface patterns. Roughly, particles covering the surface should be visible in images when their depth
is greater than the reflectivity wavelength (on the order of nanometers). All figures shown use a minimum value for
the deposition profile of 1 nm/year. Looking at Figures \ref{fig:comp23_95} and \ref{fig:zenith}, one can faintly notice
the deposition pattern of Jet 95 in the collective deposition pattern, contributing
on the order of $1$ nm per year in particle deposition. Thus, for certain deposition contributions from Jet 95, particularly
areas that do not overlap with the deposition of other jets, to be visible in surface images, Jet 95 would have to be
continuously active for approximately one year or longer. {Note that, depending on plume parameters, in particular the
average particle deposition density as well as potential variations in speed distribution parameters, this length of time may
vary. However, a rigorous comparison extending deposition rates to explicit IR/UV ratios requires significant modeling
outside the scope of this paper, which is left for future work.} 

Although the plume itself has been active for much longer than one year, there is evidence for temporal variability of plume
emissions in several forms. There has long been speculation and confirmation of tidal stresses along Enceladus' orbit
modulating the emissions \citep{Hansen:2008ds,Hedman:2013gf,Hurford:2012bt,Hurford:2007ea,Nimmo:2014ks},
as well as evidence of long-term change in emission rates between 2005--2015 \citep{Ingersoll:2017fz}. 
Recent images show prominent jets in the plume suddenly turning on or off in successive images,
while the curtain-emissions remain relatively constant \citep{variability}. It is plausible that
highly-tilted jet sources do not stay active for long periods of time, either turning on and off or changing direction
sufficiently often that their deposition signature cannot be seen in IR/UV maps. An interesting open question is whether
highly-tilted jets are more susceptible to variability and short lifespans compared with jets near orthogonal to the surface.

\section{Conclusions}\label{sec:conc}

This work provides the first detailed look at surface deposition from the Enceladus plume, providing simulated impact
and deposition data for emissions proposed in \citet{Spitale:2007ks,Porco:2014bk} and \citet{Spitale:2015ka}, for 
particle sizes between $0.6-15\mu$m. The main structure of deposition patterns is shown to be stable with respect to
variations in model parameters, total plume mass production, and size distribution slope, and deposition
patterns are consistent with IR/UV images of Enceladus' surface. Images are used as a reference to compare
simulated deposition patterns for the jets proposed in \citet{Porco:2014bk} and a curtain as proposed in \citet{Spitale:2015ka}.

The deposition pattern resulting from highly tilted jets is not consistent with the pattern seen in color maps of Enceladus' surface.
This likely indicates that highly tilted jets are not active long enough to contribute visible particle deposition patterns on the surface.
Due to the higher velocity of discrete jets compared with curtain emissions, more of these particles are likely to escape 
Enceladus' gravity, which reduces their expected contribution to surface deposition. Nevertheless, results here indicate that,
in a long-term average, most emissions are directed approximately orthogonal to the surface. 

There remain open questions on the Enceladus plume to which surface deposition may provide insight. Estimates on the
depth of resurfaced particles would allow for estimates on the plume's lifetime, based on resurfacing rates presented here.
Surface deposits also indicate where plumes have been active, and whether there have been emissions from other areas
on Enecladus' surface. Finally, reproducing surface patterns provides validation for models of plume particle dynamics
\citep{Kempf:2010kx,mass,Schmidt:2008cf,Southworth:2015im} and insight into the structure of plume emissions at the
interface between subsurface vents and particle ejection \citep{Porco:2014bk,Spitale:2007ks,Spitale:2015ka}.

\section*{Appendix: Particle flux and surface deposition}\label{sec:surface}

In considering resurfacing of Enceladus from plume particles, we are interested in two parameters: the particle collision flux in
particles/sec/m$^2$ and the rate of particle deposition in mm/year. {In this work, we simulate many millions of plume particles
and track where they land on Enceladus' surface, counting impacts in $1^\circ$ longitude $\times 1^\circ$ latitude bins.
Dividing by the total number of particles simulated for a given particle size, we arrive at a normalized impact-rate profile, $\hat{R}_{imp}(\lambda,\phi,r_i)$, for longitude $\lambda$ and latitude $\phi$,
which gives the expected number of particle impacts in some area on the surface, $(\lambda,\phi)\times(\lambda+1^\circ,\phi+1^\circ)$, based
on a single ejected particle of size $r_i$. Normalizing $R$ by the surface area of $(\lambda,\phi)\times(\lambda+1^\circ,\phi+1^\circ)$
gives a one-particle flux profile, $F(\lambda,\phi,r_i)$, which is exactly what we build from simulation
data and store in $180\times 360$ arrays. Each array is for a fixed particle radius and plume source location, and each element
corresponds to the simulated (constant) one-particle flux over a $1^\circ$ longitude $\times$ $1^\circ$ latitude area on the
moon's surface. Equations \eqref{eq:imp2} and \eqref{eq:h2} show how to use $F(\lambda,\phi,r_i)$ to compute an impact flux and
surface deposition rate, given a mass production rate, particle density, and size distribution.}

{Next, we present a formal derivation of the surface deposition rate based on a power-law size distribution. Define $p_{size}(r) = Cr^{-\alpha}$
as the power-law particle size distribution, where $C$ is chosen such that $\int_{r_{min}}^{r_{max}} p_{size}(r)\textnormal{d}r = 1$, 
$r_{min}$ is the minimum particle radius, $r_{max}$ is the maximum particle radius, and $\alpha > 1$ the size-distribution slope.}\footnote{
Note that for a well-defined size distribution and average plume particle mass, we must choose some minimum
particle radius, $r_{\textnormal{min}} > 0$ and, if $\alpha \leq 4$, some maximum radius $r_{\textnormal{max}} < \infty$.
The minimum size particle is largely based on
the mechanical origin of the particle, of which we are interested in frozen ice grains from the subsurface ocean. A separate
population of nano grains likely result from supersonic bursts through Laval nozzles in the fractures, and corresponding
to a different size distribution. A maximum size is necessary to bound the average mass of particles, but also makes sense
physically because ejecta particle size is at least limited by the channel width of fractures from which particles are emitted
(and likely much smaller).} Assuming spherical particles, the average volume of a plume particle is given by
\begin{align} \label{eq:vav}
V_{av} = \int_{r_{min}}^{r_{max}} \frac{4}{3}\pi r^3 p_{size}(r)\textnormal{d}r, \end{align}
and average mass $M_{av} = \rho V_{av}$, for average particle density $\rho$. Note that in the case of a power law, $r_{min}\neq 0$
and $r_{max}\neq \infty$ must be fixed for $V_{av}$ to be well-defined. Now consider the particle
impact rate, $R_{imp}(\lambda,\phi)$, as a function of surface location, latitude, and longitude. Let $M^+$ denote
the plume mass production in kg/sec and $N^+ = \frac{M^+}{M_{av}}$ the expected plume production rate in particles/sec.
Impact rate can then be written as $R_{imp}(\lambda,\phi) = N^+\hat{R}_{imp}(\lambda,\phi)$, where $\hat{R}_{imp}(\lambda,\phi)$ is the normalized
contribution of a single plume particle to the impact rate at location $(\lambda,\phi)$. Impact rate can be obtained by integrating
the normalized impact rate over the size distribution as a function of particle radius $r$:
\begin{align}
R_{imp}(\lambda,\phi) & = \frac{M^+}{\rho V_{av}} \int_{r_{min}}^{r_{max}} \hat{R}_{imp}(\lambda,\phi,r) p_{size}(r)\textnormal{ d}r.\label{eq:imp}
\end{align}

Now, given $R_{imp}(\lambda,\phi)$ expressed in geographical coordinates, suppose we want the expected deposition height of plume
particles covering some area $S(\lambda_0\leq\lambda\leq\lambda_1, \phi_0\leq\phi\leq\phi_1)$ per second.
The expected total volume of particles per second is given by the product of the average volume of a plume particle \textit{impacting
in $S$}, $V_{av,S}$ (generally not equal to $V_{av}$), with the expected number of particle impacts in area $S$ per second,
$n_S=\iint_S R_{imp} \textnormal{ d}S$. To compute the average volume of impacting particles at a given location $(\lambda,\phi)$, we define 
the normalized size distribution of particles impacting at $(\lambda,\phi)$ as $p_{imp}(\lambda,\phi,r):= \frac{\hat{R}_{imp}(\lambda,\phi,r) p_{size}(r)}
{\int_{r_{min}}^{r_{max}} \hat{R}_{imp}(\lambda,\phi,r) p_{size}(r)\textnormal{ d}r}$. Then, averaging over $S$, 
\begin{align}
V_{av,S} & = \frac{ \int_{\lambda_0}^{\lambda_1}\int_{\phi_0}^{\phi_1}\int_{r_{min}}^{r_{max}} \frac{4}{3}\pi r^3 p_{imp}(\lambda,\phi,r)\cos(\phi)\textnormal{ d}r\textnormal{d}\lambda\textnormal{d}\phi}{ \int_{\lambda_0}^{\lambda_1}\int_{\phi_0}^{\phi_1} \cos(\phi) \textnormal{d}\lambda\textnormal{d}\phi} \label{eq:vol_imp}
\end{align}
Let $R_E = 249.1$ km be Enceladus' radius. Volume of particles per second in $S$ is then given by:
\begin{align}
V_S & = V_{av,S}n_S \nonumber\\
& = V_{av,S} \cdot \int_{\lambda_0}^{\lambda_1}\int_{\phi_0}^{\phi_1}
	R_E^2R_{imp}(\lambda,\phi) \cos(\phi)\textnormal{ d}\lambda\textnormal{d}\phi. \label{eq:vol}
\end{align}
The depth or height of surface deposition if we assume perfect compaction
of particles is given by $h$ such that the volume of particles (Equation \ref{eq:vol}) is equal to the volume of $S$ integrated to height $h$:
\begin{align}
\int_{R_E}^{R_E+h}\int_{\lambda_0}^{\lambda_1}\int_{\phi_0}^{\phi_1} {R}^2\cos(\phi)\textnormal{ d}\lambda\textnormal{d}\phi \textnormal{d}{R}
& = h\left(R_E^2 + R_Eh + \frac{h^2}{3}\right)\int_{\lambda_0}^{\lambda_1}\int_{\phi_0}^{\phi_1} \cos(\phi)\textnormal{ d}\lambda\textnormal{d}\phi \nonumber\\
& = h\left[ R_E^2 \int_{\lambda_0}^{\lambda_1}\int_{\phi_0}^{\phi_1} \cos(\phi)\textnormal{ d}\lambda\textnormal{d}\phi + O(R_Eh + h^2) \right].\label{eq:h0}
\end{align}
Here, $h$ is expected on the order of mm or $\approx10^{-7}R_E$, which justifies dropping terms $O(R_Eh + h^2)$, and
we find that
\begin{align}
h & \approx \frac{M^+}{\rho}\cdot\frac{V_{av,S}}{V_{av}} \cdot\frac{ \int_{\lambda_0}^{\lambda_1}\int_{\phi_0}^{\phi_1}  \int_{r_{min}}^{r_{max}} \hat{R}_{imp}(\lambda,\phi,r) p_{size}(r)\cos(\phi)
	\textnormal{ d}r\textnormal{d}\lambda\textnormal{d}\phi}{\int_{\lambda_0}^{\lambda_1}\int_{\phi_0}^{\phi_1} \cos(\phi)\textnormal{ d}
	\lambda\textnormal{d}\phi}.\label{eq:h}
\end{align}
Note that dropping terms $O(R_Eh + h^2)$ is equivalent to estimating the total volume as the surface area times height.
To estimate $h$ over the moon's surface, we consider a mesh on the moon's surface of $1^\circ$ longitude $\times$ $1^\circ$ latitude
cells and approximate Equation \ref{eq:h} for each cell. Cells are sufficiently small that we assume $\hat{R}_{imp}$ to be constant over
each cell, which we normalize by the cell-size for a fixed one-particle flux, ${F}_{imp(\lambda,\phi)}$, with units $1/$m$^2$. Average volume
of impacting particles (Equation \ref{eq:vol_imp}) reduces to 
\begin{align*}
V_{av,S} & = \frac{\int_{r_{min}}^{r_{max}} \frac{4}{3}\pi r^3F(\lambda,\phi,r) p_{size}(r)\textnormal{ d}r}
	{\int_{r_{min}}^{r_{max}} F(\lambda,\phi,r) p_{size}(r)\textnormal{ d}r},
\end{align*}
and we can then separate integrals in Equation \ref{eq:h} to get:
\begin{align}
h(\lambda,\phi) & \approx \frac{M^+}{\rho}\cdot\frac{V_{av,S}}{V_{av}}\cdot \frac{ \int_{r_{min}}^{r_{max}} F(\lambda,\phi,r) p_{size}(r)\textnormal{d}r\cdot \int_{\lambda_0}^{\lambda_1}\int_{\phi_0}^{\phi_1}\cos(\phi)
	\textnormal{ d}  \lambda\textnormal{d}\phi}{ \int_{\lambda_0}^{\lambda_1}\int_{\phi_0}^{\phi_1} \cos(\phi)\textnormal{ d}
	\lambda\textnormal{d}\phi} \nonumber\\
& = \frac{M^+}{\rho V_{av}} \int_{r_{min}}^{r_{max}} \frac{4}{3}\pi r^3F(\lambda,\phi,r) p_{size}(r)\textnormal{ d}r. \label{eq:h3}
\end{align}
Notice that the (approximate) total volume (Equation \ref{eq:h3}) takes a similar form to the impact rate (Equation \ref{eq:imp}),
but now we are integrating over particle volume, $\frac{4}{3}\pi r^3\textnormal{d}r$. Each can be approximated using some
quadrature method with sample particle sizes $\{r_0,...,r_k\}$ and data $\{F_{(\lambda,\phi)}(r_i)\}_{i=0}^k$:
\begin{align}
R_{imp}(\lambda,\phi) & \approx \frac{M^+}{\rho V_{av}}\sum_i F_{(\lambda,\phi)}(r_i)p_{size}(r_i)w_i,\label{eq:imp2} \\
h(\lambda,\phi) & \approx \frac{4\pi M^+}{3\rho V_{av}}\sum_i r_i^3 F_{(\lambda,\phi)}(r_i)p_{size}(r_i)w_i,\label{eq:h2}
\end{align}
for quadrature weights $\{w_i\}$. {Weights for a trapezoid method, which are given in $\mu$m and based on averages of the
discrete particle sizes simulated,
are used to approximate \eqref{eq:imp2} and \eqref{eq:h2}.} Although more accurate methods could be used for quadrature as well as
higher resolution (non-constant) estimates of $\hat{R}_{imp}$, the underlying physical model is not sufficiently resolved to warrant such accuracy. 

\section{Acknowledgements} 
This research was conducted with Government support under and awarded by DoD, Air Force Office of Scientific Research,
National Defense Science and Engineering Graduate (NDSEG) Fellowship, 32 CFR 168a. This work was supported by NASA/JPL
under contract number 1503012. This work utilized the Janus supercomputer, which is supported by the National Science
Foundation (award number CNS-0821794) and the University of Colorado Boulder. The Janus supercomputer is a joint effort
of the University of Colorado Boulder, the University of Colorado Denver and the National Center for Atmospheric Research. 

\section*{References}
\small
\bibliography{sources.bib}

\begin{thebibliography}{44}
\expandafter\ifx\csname natexlab\endcsname\relax\def\natexlab#1{#1}\fi
\providecommand{\url}[1]{\texttt{#1}}
\providecommand{\href}[2]{#2}
\providecommand{\path}[1]{#1}
\providecommand{\DOIprefix}{doi:}
\providecommand{\ArXivprefix}{arXiv:}
\providecommand{\URLprefix}{URL: }
\providecommand{\Pubmedprefix}{pmid:}
\providecommand{\doi}[1]{\href{http://dx.doi.org/#1}{\path{#1}}}
\providecommand{\Pubmed}[1]{\href{pmid:#1}{\path{#1}}}
\providecommand{\bibinfo}[2]{#2}
\ifx\xfnm\relax \def\xfnm[#1]{\unskip,\space#1}\fi
\bibitem[{Brilliantov et~al.(2008)Brilliantov, Schmidt and
  Spahn}]{Brilliantov:2008be}
\bibinfo{author}{Brilliantov, N.}, \bibinfo{author}{Schmidt, J.},
  \bibinfo{author}{Spahn, F.}, \bibinfo{year}{2008}.
\newblock \bibinfo{title}{{Geysers of Enceladus: Quantitative analysis of
  qualitative models}}.
\newblock \bibinfo{journal}{Planetary and Space Science} \bibinfo{volume}{56},
  \bibinfo{pages}{1596--1606}.
\bibitem[{Connerney(1993)}]{Connerney:1993fp}
\bibinfo{author}{Connerney, J.E.P.}, \bibinfo{year}{1993}.
\newblock \bibinfo{title}{{Magnetic fields of the outer planets}}.
\newblock \bibinfo{journal}{Journal of Geophysical Research}
  \bibinfo{volume}{98}, \bibinfo{pages}{18659--18679}.
\bibitem[{{Di Sisto} and {Zanardi}(2016)}]{DiSisto:16a}
\bibinfo{author}{{Di Sisto}, R.P.}, \bibinfo{author}{{Zanardi}, M.},
  \bibinfo{year}{2016}.
\newblock \bibinfo{title}{{Surface ages of mid-size saturnian satellites}}.
\newblock \bibinfo{journal}{\icarus} \bibinfo{volume}{264},
  \bibinfo{pages}{90--101}.
\newblock \DOIprefix\doi{10.1016/j.icarus.2015.09.012},
  \href{http://arxiv.org/abs/1509.06773}{\tt arXiv:1509.06773}.
\bibitem[{{Dougherty} et~al.(2006){Dougherty}, {Khurana}, {Neubauer},
  {Russell}, {Saur}, {Leisner} and {Burton}}]{Dougherty:06a}
\bibinfo{author}{{Dougherty}, M.K.}, \bibinfo{author}{{Khurana}, K.K.},
  \bibinfo{author}{{Neubauer}, F.M.}, \bibinfo{author}{{Russell}, C.T.},
  \bibinfo{author}{{Saur}, J.}, \bibinfo{author}{{Leisner}, J.S.},
  \bibinfo{author}{{Burton}, M.E.}, \bibinfo{year}{2006}.
\newblock \bibinfo{title}{{Identification of a Dynamic Atmosphere at Enceladus
  with the Cassini Magnetometer}}.
\newblock \bibinfo{journal}{Science} \bibinfo{volume}{311},
  \bibinfo{pages}{1406--1409}.
\newblock \DOIprefix\doi{10.1126/science.1120985}.
\bibitem[{Gao et~al.(2016)Gao, Kopparla, Zhang and Ingersoll}]{Gao:2016}
\bibinfo{author}{Gao, P.}, \bibinfo{author}{Kopparla, P.},
  \bibinfo{author}{Zhang, X.}, \bibinfo{author}{Ingersoll, A.P.},
  \bibinfo{year}{2016}.
\newblock \bibinfo{title}{{Aggregate particles in the plumes of Enceladus}}.
\newblock \bibinfo{journal}{Icarus} .
\bibitem[{Goguen et~al.(2013)Goguen, Buratti, Brown, Clark, Nicholson, Hedman,
  Howell, Sotin, Cruikshank, Baines, Lawrence, Spencer and
  Blackburn}]{Goguen:2013ee}
\bibinfo{author}{Goguen, J.D.}, \bibinfo{author}{Buratti, B.J.},
  \bibinfo{author}{Brown, R.H.}, \bibinfo{author}{Clark, R.N.},
  \bibinfo{author}{Nicholson, P.D.}, \bibinfo{author}{Hedman, M.M.},
  \bibinfo{author}{Howell, R.R.}, \bibinfo{author}{Sotin, C.},
  \bibinfo{author}{Cruikshank, D.P.}, \bibinfo{author}{Baines, K.H.},
  \bibinfo{author}{Lawrence, K.J.}, \bibinfo{author}{Spencer, J.R.},
  \bibinfo{author}{Blackburn, D.G.}, \bibinfo{year}{2013}.
\newblock \bibinfo{title}{{The temperature and width of an active fissure on
  Enceladus measured with Cassini VIMS during the 14 April 2012 South Pole
  flyover}}.
\newblock \bibinfo{journal}{Icarus} \bibinfo{volume}{226},
  \bibinfo{pages}{1128--1137}.
\bibitem[{{Hansen} et~al.(2006){Hansen}, {Esposito}, {Stewart}, {Colwell},
  {Hendrix}, {Pryor}, {Shemansky} and {West}}]{Hansen:06a}
\bibinfo{author}{{Hansen}, C.J.}, \bibinfo{author}{{Esposito}, L.},
  \bibinfo{author}{{Stewart}, A.I.F.}, \bibinfo{author}{{Colwell}, J.},
  \bibinfo{author}{{Hendrix}, A.}, \bibinfo{author}{{Pryor}, W.},
  \bibinfo{author}{{Shemansky}, D.}, \bibinfo{author}{{West}, R.},
  \bibinfo{year}{2006}.
\newblock \bibinfo{title}{{Enceladus' Water Vapor Plume}}.
\newblock \bibinfo{journal}{Science} \bibinfo{volume}{311},
  \bibinfo{pages}{1422--1425}.
\newblock \DOIprefix\doi{10.1126/science.1121254}.
\bibitem[{Hansen et~al.(2008)Hansen, Esposito, Stewart, Meinke, Wallis,
  Colwell, Hendrix, Larsen, Pryor and Tian}]{Hansen:2008ds}
\bibinfo{author}{Hansen, C.J.}, \bibinfo{author}{Esposito, L.W.},
  \bibinfo{author}{Stewart, A.I.F.}, \bibinfo{author}{Meinke, B.},
  \bibinfo{author}{Wallis, B.}, \bibinfo{author}{Colwell, J.E.},
  \bibinfo{author}{Hendrix, A.R.}, \bibinfo{author}{Larsen, K.W.},
  \bibinfo{author}{Pryor, W.}, \bibinfo{author}{Tian, F.},
  \bibinfo{year}{2008}.
\newblock \bibinfo{title}{Water vapour jets inside the plume of gas leaving
  {E}nceladus}.
\newblock \bibinfo{journal}{Nature} \bibinfo{volume}{456},
  \bibinfo{pages}{477--479}.
\bibitem[{Hedman et~al.(2013)Hedman, Gosmeyer, Nicholson, Sotin, Brown, Clark,
  Baines, Buratti and Showalter}]{Hedman:2013gf}
\bibinfo{author}{Hedman, M.M.}, \bibinfo{author}{Gosmeyer, C.M.},
  \bibinfo{author}{Nicholson, P.D.}, \bibinfo{author}{Sotin, C.},
  \bibinfo{author}{Brown, R.H.}, \bibinfo{author}{Clark, R.N.},
  \bibinfo{author}{Baines, K.H.}, \bibinfo{author}{Buratti, B.J.},
  \bibinfo{author}{Showalter, M.R.}, \bibinfo{year}{2013}.
\newblock \bibinfo{title}{{An observed correlation between plume activity and
  tidal stresses on Enceladus}}.
\newblock \bibinfo{journal}{Nature} \bibinfo{volume}{500},
  \bibinfo{pages}{182--184}.
\bibitem[{Hedman et~al.(2009)Hedman, Nicholson, Showalter, Brown, Buratti and
  Clark}]{Hedman:2009ka}
\bibinfo{author}{Hedman, M.M.}, \bibinfo{author}{Nicholson, P.D.},
  \bibinfo{author}{Showalter, M.R.}, \bibinfo{author}{Brown, R.H.},
  \bibinfo{author}{Buratti, B.J.}, \bibinfo{author}{Clark, R.N.},
  \bibinfo{year}{2009}.
\newblock \bibinfo{title}{{Spectral Observations of the Enceladus Plume with
  Cassini-Vims}}.
\newblock \bibinfo{journal}{The Astrophysical Journal} \bibinfo{volume}{693},
  \bibinfo{pages}{1749--1762}.
\bibitem[{Helfenstein and Porco(2015)}]{Helfenstein:2015ja}
\bibinfo{author}{Helfenstein, P.}, \bibinfo{author}{Porco, C.C.},
  \bibinfo{year}{2015}.
\newblock \bibinfo{title}{{Enceladus{\textquoteright} Geysers: Relation to
  Geological Features}}.
\newblock \bibinfo{journal}{The Astronomical Journal} \bibinfo{volume}{150},
  \bibinfo{pages}{96}.
\bibitem[{Hendrix et~al.(2010)Hendrix, Hansen and Holsclaw}]{Hendrix:2010gq}
\bibinfo{author}{Hendrix, A.R.}, \bibinfo{author}{Hansen, C.J.},
  \bibinfo{author}{Holsclaw, G.M.}, \bibinfo{year}{2010}.
\newblock \bibinfo{title}{{The ultraviolet reflectance of Enceladus:
  Implications for surface composition}}.
\newblock \bibinfo{journal}{Icarus} \bibinfo{volume}{206},
  \bibinfo{pages}{608--617}.
\bibitem[{Hor{\'a}nyi(1996)}]{Horanyi:1996fl}
\bibinfo{author}{Hor{\'a}nyi, M.}, \bibinfo{year}{1996}.
\newblock \bibinfo{title}{{Charged dust dynamics in the solar system}}.
\newblock \bibinfo{journal}{Annual Review of Astronomy and Astrophysics}
  \bibinfo{volume}{34}, \bibinfo{pages}{383--418}.
\bibitem[{{Hor{\'a}nyi} et~al.(2009){Hor{\'a}nyi}, {Burns}, {Hedman}, {Jones}
  and {Kempf}}]{Horanyi:09a}
\bibinfo{author}{{Hor{\'a}nyi}, M.}, \bibinfo{author}{{Burns}, J.A.},
  \bibinfo{author}{{Hedman}, M.M.}, \bibinfo{author}{{Jones}, G.H.},
  \bibinfo{author}{{Kempf}, S.}, \bibinfo{year}{2009}.
\newblock \bibinfo{title}{{Diffuse Rings}}, in: \bibinfo{editor}{{Dougherty,
  M.~K., Esposito, L.~W., \& Krimigis, S.~M.}} (Ed.),
  \bibinfo{booktitle}{Saturn from Cassini-Huygens}.
  \bibinfo{publisher}{Springer}, pp. \bibinfo{pages}{511--536}.
\newblock \DOIprefix\doi{10.1007/978-1-4020-9217-6_16}.
\bibitem[{Howett et~al.(2011)Howett, Spencer, Pearl and Segura}]{Howett:2011gw}
\bibinfo{author}{Howett, C.J.A.}, \bibinfo{author}{Spencer, J.R.},
  \bibinfo{author}{Pearl, J.C.}, \bibinfo{author}{Segura, M.},
  \bibinfo{year}{2011}.
\newblock \bibinfo{title}{{High heat flow from Enceladus' south polar region
  measured using 10-600 cm-1 Cassini/CIRS data}}.
\newblock \bibinfo{journal}{Journal of Geophysical Research}
  \bibinfo{volume}{116}, \bibinfo{pages}{E03003}.
\bibitem[{Hurford et~al.(2007)Hurford, Helfenstein, Hoppa, Greenberg and
  Bills}]{Hurford:2007ea}
\bibinfo{author}{Hurford, T.A.}, \bibinfo{author}{Helfenstein, P.},
  \bibinfo{author}{Hoppa, G.V.}, \bibinfo{author}{Greenberg, R.},
  \bibinfo{author}{Bills, B.G.}, \bibinfo{year}{2007}.
\newblock \bibinfo{title}{{Eruptions arising from tidally controlled periodic
  openings of rifts on Enceladus}}.
\newblock \bibinfo{journal}{Nature} \bibinfo{volume}{447},
  \bibinfo{pages}{292--294}.
\bibitem[{Hurford et~al.(2012)Hurford, Helfenstein and
  Spitale}]{Hurford:2012bt}
\bibinfo{author}{Hurford, T.A.}, \bibinfo{author}{Helfenstein, P.},
  \bibinfo{author}{Spitale, J.N.}, \bibinfo{year}{2012}.
\newblock \bibinfo{title}{{Tidal control of jet eruptions on Enceladus as
  observed by Cassini ISS between 2005 and 2007}}.
\newblock \bibinfo{journal}{Icarus} \bibinfo{volume}{220},
  \bibinfo{pages}{896--903}.
\bibitem[{Ingersoll and Ewald(2017)}]{Ingersoll:2017fz}
\bibinfo{author}{Ingersoll, A.P.}, \bibinfo{author}{Ewald, S.P.},
  \bibinfo{year}{2017}.
\newblock \bibinfo{title}{{Decadal timescale variability of the Enceladus
  plumes inferred from Cassini images}}.
\newblock \bibinfo{journal}{Icarus} \bibinfo{volume}{282},
  \bibinfo{pages}{260--275}.
\bibitem[{{Jaumann} et~al.(2009){Jaumann}, {Clark}, {Nimmo}, {Hendrix},
  {Buratti}, {Denk}, {Moore}, {Schenk}, {Ostro} and {Srama}}]{Jaumann:09a}
\bibinfo{author}{{Jaumann}, R.}, \bibinfo{author}{{Clark}, R.N.},
  \bibinfo{author}{{Nimmo}, F.}, \bibinfo{author}{{Hendrix}, A.R.},
  \bibinfo{author}{{Buratti}, B.J.}, \bibinfo{author}{{Denk}, T.},
  \bibinfo{author}{{Moore}, J.M.}, \bibinfo{author}{{Schenk}, P.M.},
  \bibinfo{author}{{Ostro}, S.J.}, \bibinfo{author}{{Srama}, R.},
  \bibinfo{year}{2009}.
\newblock \bibinfo{title}{{Icy Satellites: Geological Evolution and Surface
  Processes}}.
\newblock pp. \bibinfo{pages}{637--681}.
\newblock \DOIprefix\doi{10.1007/978-1-4020-9217-6_20}.
\bibitem[{Kempf et~al.(2010)Kempf, Beckmann and Schmidt}]{Kempf:2010kx}
\bibinfo{author}{Kempf, S.}, \bibinfo{author}{Beckmann, U.},
  \bibinfo{author}{Schmidt, J.}, \bibinfo{year}{2010}.
\newblock \bibinfo{title}{{How the Enceladus dust plume feeds
  Saturn{\textquoteright}s E ring}}.
\newblock \bibinfo{journal}{Icarus} \bibinfo{volume}{206},
  \bibinfo{pages}{446--457}.
\bibitem[{Meier et~al.(2015)Meier, Motschmann, Schmidt and
  Spahn}]{Meier:2015ba}
\bibinfo{author}{Meier, P.}, \bibinfo{author}{Motschmann, U.},
  \bibinfo{author}{Schmidt, J.}, \bibinfo{author}{Spahn, F.},
  \bibinfo{year}{2015}.
\newblock \bibinfo{title}{{Modeling the total dust production of enceladus from
  stochastic charge equilibrium and simulations}}.
\newblock \bibinfo{journal}{Planetary and Space Science} \bibinfo{volume}{119},
  \bibinfo{pages}{208--221}.
\bibitem[{{Nahm} and {Kattenhorn}(2015)}]{Nahm:16a}
\bibinfo{author}{{Nahm}, A.L.}, \bibinfo{author}{{Kattenhorn}, S.A.},
  \bibinfo{year}{2015}.
\newblock \bibinfo{title}{{A unified nomenclature for tectonic structures on
  the surface of Enceladus}}.
\newblock \bibinfo{journal}{\icarus} \bibinfo{volume}{258},
  \bibinfo{pages}{67--81}.
\newblock \DOIprefix\doi{10.1016/j.icarus.2015.06.009}.
\bibitem[{Nimmo et~al.(2014)Nimmo, Porco and Mitchell}]{Nimmo:2014ks}
\bibinfo{author}{Nimmo, F.}, \bibinfo{author}{Porco, C.},
  \bibinfo{author}{Mitchell, C.}, \bibinfo{year}{2014}.
\newblock \bibinfo{title}{{Tidally Modulated Eruptions on Enceladus: Cassini
  ISS Observations and Models}}.
\newblock \bibinfo{journal}{The Astronomical Journal} \bibinfo{volume}{148},
  \bibinfo{pages}{46}.
\bibitem[{Porco et~al.(2014)Porco, DiNino and Nimmo}]{Porco:2014bk}
\bibinfo{author}{Porco, C.C.}, \bibinfo{author}{DiNino, D.},
  \bibinfo{author}{Nimmo, F.}, \bibinfo{year}{2014}.
\newblock \bibinfo{title}{{How the Geysers, Tidal Stresses, and Thermal
  Emission across the South Polar Terrain of Enceladus are Related}}.
\newblock \bibinfo{journal}{The Astronomical Journal} \bibinfo{volume}{148},
  \bibinfo{pages}{45}.
\bibitem[{Porco et~al.(2017)Porco, Dones and Mitchell}]{Porco:2017kh}
\bibinfo{author}{Porco, C.C.}, \bibinfo{author}{Dones, L.},
  \bibinfo{author}{Mitchell, C.}, \bibinfo{year}{2017}.
\newblock \bibinfo{title}{{Could It Be Snowing Microbes on Enceladus? Assessing
  Conditions in Its Plume and Implications for Future Missions}}.
\newblock \bibinfo{journal}{Astrobiology} , \bibinfo{pages}{ast.2017.1665--26}.
\bibitem[{Porco et~al.(2006)Porco, Helfenstein, Thomas, Ingersoll, Wisdom,
  West, Neukum, Denk, Wagner, Roatsch, Kieffer, Turtle, McEwen, Johnson,
  Rathbun, Veverka, Wilson, Perry, Spitale, Brahic, Burns, Delgenio, Dones,
  Murray and Squyres}]{Porco:2006ir}
\bibinfo{author}{Porco, C.C.}, \bibinfo{author}{Helfenstein, P.},
  \bibinfo{author}{Thomas, P.C.}, \bibinfo{author}{Ingersoll, A.P.},
  \bibinfo{author}{Wisdom, J.}, \bibinfo{author}{West, R.A.},
  \bibinfo{author}{Neukum, G.}, \bibinfo{author}{Denk, T.},
  \bibinfo{author}{Wagner, R.}, \bibinfo{author}{Roatsch, T.},
  \bibinfo{author}{Kieffer, S.W.}, \bibinfo{author}{Turtle, E.},
  \bibinfo{author}{McEwen, A.}, \bibinfo{author}{Johnson, T.V.},
  \bibinfo{author}{Rathbun, J.}, \bibinfo{author}{Veverka, J.},
  \bibinfo{author}{Wilson, D.}, \bibinfo{author}{Perry, J.},
  \bibinfo{author}{Spitale, J.N.}, \bibinfo{author}{Brahic, A.},
  \bibinfo{author}{Burns, J.A.}, \bibinfo{author}{Delgenio, A.},
  \bibinfo{author}{Dones, L.}, \bibinfo{author}{Murray, C.D.},
  \bibinfo{author}{Squyres, S.}, \bibinfo{year}{2006}.
\newblock \bibinfo{title}{{Cassini observes the active South Pole of
  Enceladus}}.
\newblock \bibinfo{journal}{Science} \bibinfo{volume}{311},
  \bibinfo{pages}{1393--1401}.
\bibitem[{Postberg et~al.(2009)Postberg, Kempf, Schmidt, Brilliantov, Beinsen,
  Abel, Buck and Srama}]{Postberg:2009jc}
\bibinfo{author}{Postberg, F.}, \bibinfo{author}{Kempf, S.},
  \bibinfo{author}{Schmidt, J.}, \bibinfo{author}{Brilliantov, N.},
  \bibinfo{author}{Beinsen, A.}, \bibinfo{author}{Abel, B.},
  \bibinfo{author}{Buck, U.}, \bibinfo{author}{Srama, R.},
  \bibinfo{year}{2009}.
\newblock \bibinfo{title}{{Sodium salts in E-ring ice grains from an ocean
  below the surface of Enceladus}}.
\newblock \bibinfo{journal}{Nature} \bibinfo{volume}{459},
  \bibinfo{pages}{1098--1101}.
\bibitem[{Postberg et~al.(2011)Postberg, Schmidt, Hillier, Kempf and
  Srama}]{Postberg:2011jx}
\bibinfo{author}{Postberg, F.}, \bibinfo{author}{Schmidt, J.},
  \bibinfo{author}{Hillier, J.K.}, \bibinfo{author}{Kempf, S.},
  \bibinfo{author}{Srama, R.}, \bibinfo{year}{2011}.
\newblock \bibinfo{title}{{A salt-water reservoir as the source of a
  compositionally stratified plume on Enceladus}}.
\newblock \bibinfo{journal}{Nature} \bibinfo{volume}{474},
  \bibinfo{pages}{620--622}.
\bibitem[{Schenk et~al.(2011)Schenk, Hamilton, Johnson, McKinnon, Paranicas,
  Schmidt and Showalter}]{Schenk:2011bv}
\bibinfo{author}{Schenk, P.}, \bibinfo{author}{Hamilton, D.P.},
  \bibinfo{author}{Johnson, R.E.}, \bibinfo{author}{McKinnon, W.B.},
  \bibinfo{author}{Paranicas, C.}, \bibinfo{author}{Schmidt, J.},
  \bibinfo{author}{Showalter, M.R.}, \bibinfo{year}{2011}.
\newblock \bibinfo{title}{{Plasma, plumes and rings: Saturn system dynamics as
  recorded in global color patterns on its midsize icy satellites}}.
\newblock \bibinfo{journal}{Icarus} \bibinfo{volume}{211},
  \bibinfo{pages}{740--757}.
\bibitem[{Schmidt et~al.(2008)Schmidt, Brilliantov, Spahn and
  Kempf}]{Schmidt:2008cf}
\bibinfo{author}{Schmidt, J.}, \bibinfo{author}{Brilliantov, N.},
  \bibinfo{author}{Spahn, F.}, \bibinfo{author}{Kempf, S.},
  \bibinfo{year}{2008}.
\newblock \bibinfo{title}{{Slow dust in Enceladus' plume from condensation and
  wall collisions in tiger stripe fractures}}.
\newblock \bibinfo{journal}{Nature} \bibinfo{volume}{451},
  \bibinfo{pages}{685--688}.
\bibitem[{Scipioni et~al.(2017)Scipioni, Schenk, Tosi, D'Aversa, Clark, Combe
  and Ore}]{Scipioni:2017is}
\bibinfo{author}{Scipioni, F.}, \bibinfo{author}{Schenk, P.},
  \bibinfo{author}{Tosi, F.}, \bibinfo{author}{D'Aversa, E.},
  \bibinfo{author}{Clark, R.N.}, \bibinfo{author}{Combe, J.P.},
  \bibinfo{author}{Ore, C.M.D.}, \bibinfo{year}{2017}.
\newblock \bibinfo{title}{{Deciphering sub-micron ice particles on Enceladus
  surface}}.
\newblock \bibinfo{journal}{Icarus} \bibinfo{volume}{290},
  \bibinfo{pages}{183--200}.
\bibitem[{Simon et~al.(2011)Simon, Saur, Kriegel, Neubauer, Motschmann and
  Dougherty}]{Simon:2011ef}
\bibinfo{author}{Simon, S.}, \bibinfo{author}{Saur, J.},
  \bibinfo{author}{Kriegel, H.}, \bibinfo{author}{Neubauer, F.M.},
  \bibinfo{author}{Motschmann, U.}, \bibinfo{author}{Dougherty, M.K.},
  \bibinfo{year}{2011}.
\newblock \bibinfo{title}{{Influence of negatively charged plume grains and
  hemisphere coupling currents on the structure of Enceladus' Alfv{\'e}n wings:
  Analytical modeling of Cassini magnetometer observations}}.
\newblock \bibinfo{journal}{Journal of Geophysical Research}
  \bibinfo{volume}{116}, \bibinfo{pages}{4221}.
\bibitem[{Southworth et~al.(2015)Southworth, Kempf and
  Schmidt}]{Southworth:2015im}
\bibinfo{author}{Southworth, B.S.}, \bibinfo{author}{Kempf, S.},
  \bibinfo{author}{Schmidt, J.}, \bibinfo{year}{2015}.
\newblock \bibinfo{title}{{Modeling Europa's dust plumes}}.
\newblock \bibinfo{journal}{Geophysical Research Letters} \bibinfo{volume}{42}.
\bibitem[{Southworth et~al.(2018)Southworth, Kempf, Schmidt, Postberg, Economou
  and Moragas-Klostermeyer}]{mass}
\bibinfo{author}{Southworth, B.S.}, \bibinfo{author}{Kempf, S.},
  \bibinfo{author}{Schmidt, J.}, \bibinfo{author}{Postberg, F.},
  \bibinfo{author}{Economou, T.}, \bibinfo{author}{Moragas-Klostermeyer, G.},
  \bibinfo{year}{2018}.
\newblock \bibinfo{title}{{CDA} encounters the {E}nceladus plume: mass
  production and particle distributions}.
\newblock \bibinfo{journal}{in preparation} .
\bibitem[{{Spahn} et~al.(2006){Spahn}, {Albers}, {H{\"o}rning}, {Kempf},
  {Krivov}, {Makuch}, {Schmidt}, {Sei{\ss}} and {Srem{\v
  c}evi{\'c}}}]{Spahn:06b}
\bibinfo{author}{{Spahn}, F.}, \bibinfo{author}{{Albers}, N.},
  \bibinfo{author}{{H{\"o}rning}, M.}, \bibinfo{author}{{Kempf}, S.},
  \bibinfo{author}{{Krivov}, A.V.}, \bibinfo{author}{{Makuch}, M.},
  \bibinfo{author}{{Schmidt}, J.}, \bibinfo{author}{{Sei{\ss}}, M.},
  \bibinfo{author}{{Srem{\v c}evi{\'c}}, M.}, \bibinfo{year}{2006}.
\newblock \bibinfo{title}{{E ring dust sources: Implications from Cassini's
  dust measurements}}.
\newblock \bibinfo{journal}{Planet. Space Sci.} \bibinfo{volume}{54},
  \bibinfo{pages}{1024--1032}.
\newblock \DOIprefix\doi{10.1016/j.pss.2006.05.022}.
\bibitem[{Spahn et~al.(2006)Spahn, Schmidt, Albers, Horning, Makuch, Seiss,
  Kempf, Srama, Dikarev, Helfert, Moragas-Klostermeyer, Krivov, Srem{\v
  c}evi{\'c}, Tuzzolino, Economou and Gr{\"u}n}]{Spahn:2006im}
\bibinfo{author}{Spahn, F.}, \bibinfo{author}{Schmidt, J.},
  \bibinfo{author}{Albers, N.}, \bibinfo{author}{Horning, M.},
  \bibinfo{author}{Makuch, M.}, \bibinfo{author}{Seiss, M.},
  \bibinfo{author}{Kempf, S.}, \bibinfo{author}{Srama, R.},
  \bibinfo{author}{Dikarev, V.}, \bibinfo{author}{Helfert, S.},
  \bibinfo{author}{Moragas-Klostermeyer, G.}, \bibinfo{author}{Krivov, A.V.},
  \bibinfo{author}{Srem{\v c}evi{\'c}, M.}, \bibinfo{author}{Tuzzolino, A.J.},
  \bibinfo{author}{Economou, T.}, \bibinfo{author}{Gr{\"u}n, E.},
  \bibinfo{year}{2006}.
\newblock \bibinfo{title}{{Cassini dust measurements at Enceladus and
  implications for the origin of the E ring}}.
\newblock \bibinfo{journal}{Science} \bibinfo{volume}{311},
  \bibinfo{pages}{1416--1418}.
\bibitem[{Spencer et~al.(2006)Spencer, Pearl, Segura, Flasar, Mamoutkine,
  Romani, Buratti, Hendrix, Spilker and Lopes}]{Spencer:2006gt}
\bibinfo{author}{Spencer, J.R.}, \bibinfo{author}{Pearl, J.C.},
  \bibinfo{author}{Segura, M.}, \bibinfo{author}{Flasar, F.M.},
  \bibinfo{author}{Mamoutkine, A.}, \bibinfo{author}{Romani, P.},
  \bibinfo{author}{Buratti, B.J.}, \bibinfo{author}{Hendrix, A.R.},
  \bibinfo{author}{Spilker, L.J.}, \bibinfo{author}{Lopes, R.M.C.},
  \bibinfo{year}{2006}.
\newblock \bibinfo{title}{{Cassini encounters Enceladus: Background and the
  discovery of a south polar hot spot}}.
\newblock \bibinfo{journal}{Science} \bibinfo{volume}{311},
  \bibinfo{pages}{1401--1405}.
\bibitem[{Spitale et~al.(2015)Spitale, Hurford, Rhoden, Berkson and
  Platts}]{Spitale:2015ka}
\bibinfo{author}{Spitale, J.N.}, \bibinfo{author}{Hurford, T.A.},
  \bibinfo{author}{Rhoden, A.R.}, \bibinfo{author}{Berkson, E.E.},
  \bibinfo{author}{Platts, S.S.}, \bibinfo{year}{2015}.
\newblock \bibinfo{title}{{Curtain eruptions from Enceladus{\textquoteright}
  south-polar terrain}}.
\newblock \bibinfo{journal}{Nature} \bibinfo{volume}{521},
  \bibinfo{pages}{57--60}.
\bibitem[{Spitale and Porco(2007)}]{Spitale:2007ks}
\bibinfo{author}{Spitale, J.N.}, \bibinfo{author}{Porco, C.C.},
  \bibinfo{year}{2007}.
\newblock \bibinfo{title}{{Association of the jets of Enceladus with the
  warmest regions on its south-polar fractures}}.
\newblock \bibinfo{journal}{Nature} \bibinfo{volume}{449},
  \bibinfo{pages}{695--697}.
\bibitem[{Spitale and Southworth(2017)}]{variability}
\bibinfo{author}{Spitale, J.N.}, \bibinfo{author}{Southworth, B.S.},
  \bibinfo{year}{2017}.
\newblock \bibinfo{title}{Short-term varaibility in the enceladus plume}.
\newblock \bibinfo{journal}{submitted} .
\bibitem[{Teolis et~al.(2017)Teolis, Perry, Hansen, Waite, Porco, Spencer and
  Howett}]{teolis2017enceladus}
\bibinfo{author}{Teolis, B.D.}, \bibinfo{author}{Perry, M.E.},
  \bibinfo{author}{Hansen, C.J.}, \bibinfo{author}{Waite, J.H.},
  \bibinfo{author}{Porco, C.C.}, \bibinfo{author}{Spencer, J.R.},
  \bibinfo{author}{Howett, C.J.}, \bibinfo{year}{2017}.
\newblock \bibinfo{title}{Enceladus plume structure and time variability:
  Comparison of cassini observations}.
\newblock \bibinfo{journal}{Astrobiology} \bibinfo{volume}{17},
  \bibinfo{pages}{926--940}.
\bibitem[{{The HDF Group}(2000-2010)}]{hdf5}
\bibinfo{author}{{The HDF Group}}, \bibinfo{year}{2000-2010}.
\newblock \bibinfo{title}{{Hierarchical data format version 5}}.
\newblock \URLprefix \url{http://www.hdfgroup.org/HDF5}.
\bibitem[{{Waite} et~al.(2009){Waite}, {Lewis}, {Magee}, {Lunine}, {McKinnon},
  {Glein}, {Mousis}, {Young}, {Brockwell}, {Westlake}, {Nguyen}, {Teolis},
  {Niemann}, {McNutt}, {Perry} and {Ip}}]{Waite:09a}
\bibinfo{author}{{Waite}, Jr., J.H.}, \bibinfo{author}{{Lewis}, W.S.},
  \bibinfo{author}{{Magee}, B.A.}, \bibinfo{author}{{Lunine}, J.I.},
  \bibinfo{author}{{McKinnon}, W.B.}, \bibinfo{author}{{Glein}, C.R.},
  \bibinfo{author}{{Mousis}, O.}, \bibinfo{author}{{Young}, D.T.},
  \bibinfo{author}{{Brockwell}, T.}, \bibinfo{author}{{Westlake}, J.},
  \bibinfo{author}{{Nguyen}, M.}, \bibinfo{author}{{Teolis}, B.D.},
  \bibinfo{author}{{Niemann}, H.B.}, \bibinfo{author}{{McNutt}, R.L.},
  \bibinfo{author}{{Perry}, M.}, \bibinfo{author}{{Ip}, W.},
  \bibinfo{year}{2009}.
\newblock \bibinfo{title}{Liquid water on enceladus from observations of
  ammonia and {$^{40}$}ar in the plume}.
\newblock \bibinfo{journal}{\nat} \bibinfo{volume}{460},
  \bibinfo{pages}{487--490}.
\newblock \DOIprefix\doi{10.1038/nature08153}.
\bibitem[{Yeoh et~al.(2015)Yeoh, Chapman, Goldstein, Varghese and
  Trafton}]{Yeoh:2015hg}
\bibinfo{author}{Yeoh, S.K.}, \bibinfo{author}{Chapman, T.A.},
  \bibinfo{author}{Goldstein, D.B.}, \bibinfo{author}{Varghese, P.L.},
  \bibinfo{author}{Trafton, L.M.}, \bibinfo{year}{2015}.
\newblock \bibinfo{title}{{On understanding the physics of the Enceladus south
  polar plume via numerical simulation}}.
\newblock \bibinfo{journal}{Icarus} \bibinfo{volume}{253},
  \bibinfo{pages}{205--222}.

\end{thebibliography}

\end{document}